\documentclass[epjST]{svjour}
\usepackage{hyperref}
\usepackage{graphicx,graphics,epsfig}
\usepackage{times,bm,bbm,amssymb,amsmath,amsfonts,mathrsfs}
\usepackage{subfigure}
\usepackage{dsfont}
\usepackage{braket}
\usepackage{upgreek}
\usepackage{cite}

\newcommand{\besAlbash} {\begin{subequations}}
\newcommand{\eesAlbash} {\end{subequations}}
\newcommand{\beqAlbash} {\begin{equation}}
\newcommand{\eeqAlbash} {\end{equation}}

\begin{document}
\title{Reexamining classical and quantum models for the D-Wave One processor}
\subtitle{The role of excited states and ground state degeneracy}
\author{Tameem Albash\inst{1}\inst{2} \and Troels F. R\o nnow\inst{3}  \and Matthias Troyer\inst{3} \and Daniel A. Lidar\inst{1}\inst{2}\inst{4}\inst{5}\fnmsep\thanks{\email{lidar@usc.edu}}}
\institute{Department of Physics and Astronomy, University of Southern California, Los Angeles, California 90089, USA \and Center for Quantum Information Science \& Technology, University of Southern California, Los Angeles, California 90089, USA \and Theoretische Physik, ETH Zurich, 8093 Zurich, Switzerland \and Department of Chemistry, University of Southern California, Los Angeles, California 90089, USA \and Department of Electrical Engineering, University of Southern California, Los Angeles, California 90089, USA}
\abstract{
We revisit the evidence for quantum annealing in the D-Wave One device (DW1) based on the study of random Ising instances. Using the probability distributions of finding the ground states of such instances, previous work 
found agreement with both simulated quantum annealing (SQA) and a classical rotor model. Thus the DW1 ground state success probabilities are consistent with both models, and a different measure is needed to distinguish the data and the models. Here we consider measures that account for ground state degeneracy and the distributions of excited states, and present evidence that for these new measures neither SQA nor the classical rotor model correlate perfectly with the DW1 experiments. We thus provide evidence that SQA and the classical rotor model, both of which are classically efficient algorithms, do not satisfactorily explain all the DW1 data. A complete model for the DW1  remains an open problem. Using the same criteria we find that, on the other hand, SQA and the classical rotor model correlate closely with each other. To explain this we show that the rotor model can be derived as the semiclassical limit of the spin-coherent states path integral. We also find differences in which set of  ground states is found by each method, though this feature is sensitive to calibration errors of the DW1 device and to simulation parameters.} 
%
\maketitle
\section{Introduction}
\label{Albash-intro}

\begin{figure}[t]
 \centering
 \includegraphics[width=\textwidth]{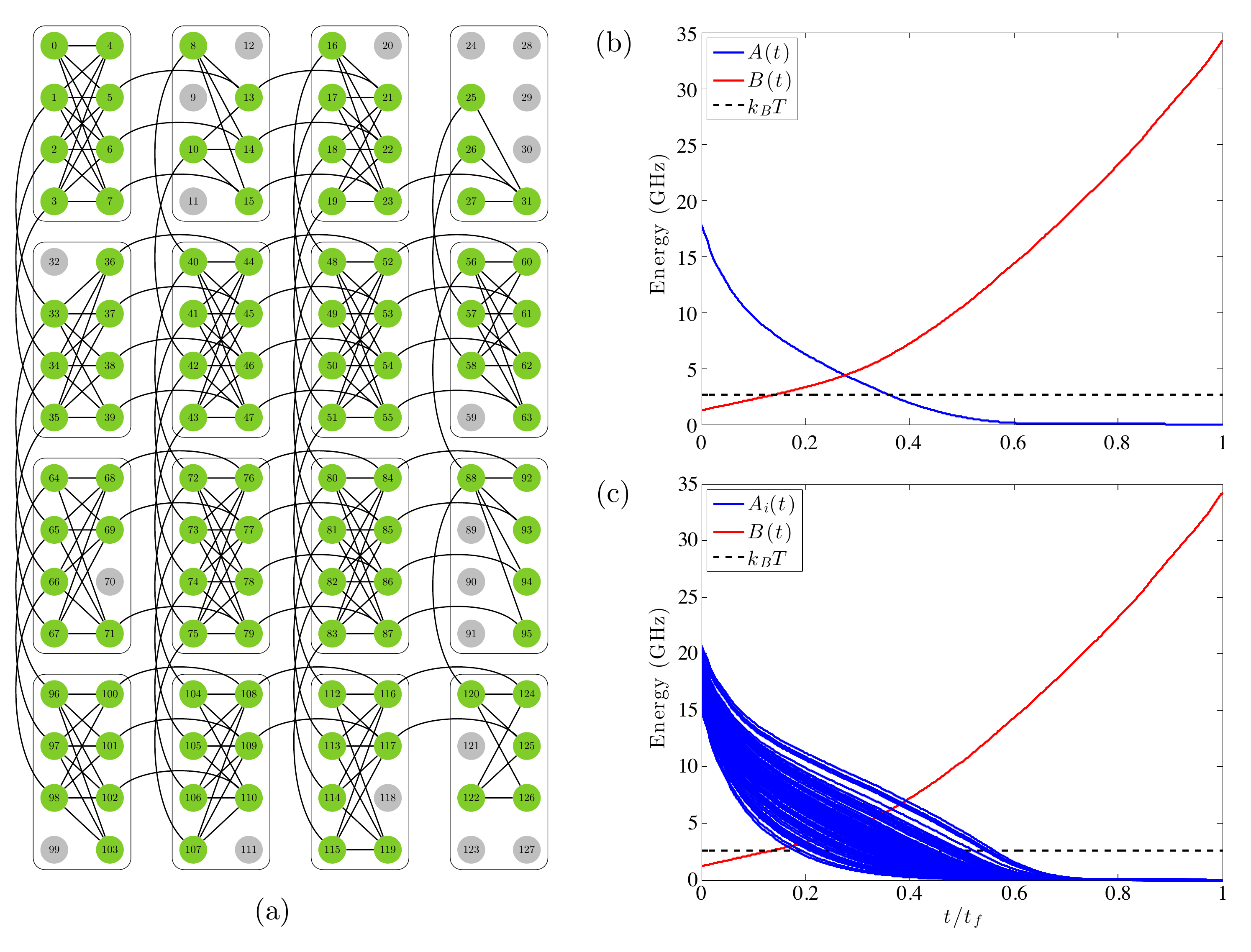}
	\caption{(a) Chimera graph of the DW1. Black lines indicate couplings $J_{ij}$ between pairs of functional qubits $i$ and $j$ (green circles). Grey circles indicate non-working qubits. Panel (b) shows the DW1 annealing schedule, in units of $\hbar = 1$.  In (b) the same median annealing schedules is used on all qubits. In (c) we allow for a unique transverse field annealing schedule for each qubit. We used both (b) and (c) in our SQA and SSSV simulations. The DW1 operating temperature of $2.61$GHz  ($20$mK) is indicated by the dashed horizontal line.}
\label{Albash-fig:DW1}
\end{figure}

The devices manufactured by D-Wave Systems Inc. \cite{Johnson:2010ys,Berkley:2010zr,Harris:2010kx,Bunyk:2014hb} are candidate physical implementations of the quantum annealing algorithm \cite{finnila_quantum_1994,kadowaki_quantum_1998,brooke_quantum_1999,brooke_tunable_2001,Santoro:2006hh,morita:125210,RevModPhys.80.1061,Bapst2013}. Although quantum tunneling \cite{Dwave} and entanglement \cite{DWave-entanglement} were demonstrated in D-Wave devices, whether quantum effects play a decisive role in determining the output statistics of such devices has remained a controversial question. A recent attempt to answer this question used random Ising model instances and compared the output statistics of a $108$-qubit D-Wave One ``Rainier'' (DW1) device to several algorithms \cite{q108}. The comparison focused on the probabilities of successfully finding the ground states of these random Ising instances, defined via the Hamiltonian
\beqAlbash
H_{\textrm{Ising}} =  \sum_{\{(i,j)\}\in E(G)} J_{ij} \sigma^z_i \sigma^z_j + \sum_{i \in V(G)} h_i \sigma^z_i \ ,
\eeqAlbash
where $\{J_{ij}\}$ are the Ising couplings along the edge set $E$ of the Chimera graph $G$ shown in Fig.~\ref{Albash-fig:DW1}(a), $h_i$ are local fields on the vertices $V$ of $G$, and $\sigma^z_i =\pm 1$ are the Ising spin variables. The study found a poor correlation between the success probabilities of the DW1 and classical simulated annealing \cite{kirkpatrick_optimization_1983} as well as a classical spin dynamics model \cite{Smolin,comment-SS} that is similar to the Landau-Lifshitz-Gilbert equation \cite{Gilbert:04}. At the same time it found a good correlation between the DW1 success probabilities and simulated quantum annealing (SQA, a quantum Monte Carlo algorithm which samples thermal equilibrium state distribution) \cite{sqa1,Santoro,Bapst2013}. The experiment thus rejected a pair of classical models and provided evidence that the DW1 was performing in a manner consistent with (finite temperature) quantum annealing. 


This, of course, left open the possibility of there being another ``classical'' model that is consistent with the DW1 data. Indeed, Shin, Smith, Smolin \& Vazirani (SSSV) provided just such a model \cite{SSSV}, where qubits are replaced by interacting planar rotors that follow the D-Wave annealing schedule and whose angles are updated via Metropolis moves. This model correlates as well with the DW1 success probabilities for random Ising instances as SQA.  Furthermore, the SSSV model success probabilities also correlate very strongly with the SQA success probabilities over the same data set \cite{SSSV}.  

Perhaps the most surprising aspect of the SSSV model, which performs a quantum-annealing like schedule on classical rotors, is its strong correlation with SQA over the random Ising instance data set. One possibility to explain this is that SSSV represents a mean-field description that accurately describes the output of SQA for the class of transverse-field Ising problems studied. In this sense the surprise would be diminished, since quantum systems often have phases that are captured by classical models with renormalized parameters. However, in the SSSV model the dynamics of the rotors can be interpreted as describing the evolution of coherent single qubits (via the standard mapping of a qubit to the Bloch sphere) without multi-qubit coherence, and in particular without any entanglement; the fact that the SSSV model  correlates strongly with both the DW1 and SQA would then suggest that multi-qubit quantum effects do not play a role in determining the success probability distribution for the random Ising problems studied in Ref.~\cite{q108}.

Note that SQA and SSSV are both classical algorithms that run efficiently on classical computers. SQA is, however, a classical algorithm describing a quantum mechanical model and thus explicitly accounts for entanglement and some aspects of tunneling, while SSSV simulates a classical model that  accounts for the effects of single qubit tunneling and potentially mimics SQA by absorbing the effects of entanglement and multi-qubit tunneling into a renormalization of its parameters.

These results and observations raise several questions. Is there a way to distinguish the two models from the results of experiments on the D-Wave devices? What is the relation between SQA and the SSSV model? So far the answers to these questions have remained largely elusive. In this work, we examine additional DW1 data from the same set of random Ising instances studied in Ref.~\cite{q108}, specifically excited states and ground state degeneracy, and attempt to elucidate this situation. In particular, we show that neither SQA nor the SSSV model provides a complete description of the DW1 results beyond the success probability data, that the two models are largely indistinguishable over the random Ising set of instances considered in Ref.~\cite{q108}, and that the SSSV model can be derived from the spin-coherent state path integral formulation of SQA.  

We remark that meanwhile another study examined the SSSV model in light of data from a $503$-qubit D-Wave Two ``Vesuvius'' (DW2) device, from specifically designed Ising instances on up to $20$ qubits, first proposed in Ref.~\cite{q-sig}, and found strong discrepancies on these grounds \cite{q-sig2} (overcoming objections raised in Ref.~\cite{SSSV-comment}). At the same time it found that an adiabatic quantum master equation derived in the weak system-bath coupling limit \cite{ABLZ:12-SI} agrees very well with the DW2 data on instances of up to $8$ qubits; beyond this the master equation simulations became prohibitively time-consuming. It is possible that other approaches (e.g., Ising instances with high tunneling barriers \cite{Vadim-private-comm}) will shed more light on the suitability of SQA and the SSSV model; here we choose to focus solely on the random Ising instances used in the work of Ref.~\cite{q108} as they provide a rich source of data with previously unexamined aspects.

The structure of this paper is as follows. In Section~\ref{sec:expt-models} we briefly review the D-Wave device, the SSSV model, and SQA.
In Section~\ref{sec:espd} we study one measure (introduced in Ref.~\cite{q108}) that allows us to compare the DW1, SQA and SSSV in terms of the distribution of excited state energies and success probabilities. It turns out that this measure is too coarse to distinguish the models from each other or the experiment. For this reason we introduce a different measure in Section~\ref{sec:dist}, the total variation distance between the probability distributions over excited states. This measure allows us to compare the energy spectra on an instance by instance basis.  We demonstrate that using this measure, neither SQA nor SSSV correlate perfectly with DW1, while SQA and SSSV remain well correlated with each other.  We show that this conclusion is robust to a large variation of model parameters, suggesting that although SSSV and SQA do reproduce the success probabilities of the device, they do not explain the full state spectrum observed. In Section~\ref{sec:GS} we switch our attention to a comparison of the distribution of ground states found by each method. We show that this measure is more sensitive than the ground state probabilities used in Ref.~\cite{q108}, and can be used to distinguish the DW1 from SQA and SSSV. The latter two remain strongly correlated when compared in terms of a distance measure between the ground state subspaces, but differ when compared in terms of the actual set of degenerate ground states found for a given random Ising instance. Given the overall strong agreement between SQA and SSSV we present, in Section~\ref{sec:SSSV-deriv}, a path integral derivation of the SSSV model in a certain semi-classical limit, which allows us to connect the SSSV model to a mean-field approximation of SQA. We summarize our findings and conclude in Section~\ref{sec:conc}.

\section{The experiment and the models}
\label{sec:expt-models}

The D-Wave devices and both models have been described in detail elsewhere (see, e.g., the Supplementary Information of Ref.~\cite{q108} for the DW1 and SQA, and Ref.~\cite{SSSV} for the SSSV model), so we provide only a brief description here. 

The D-Wave devices are designed to implement quantum annealing (QA) by evolving a system of $N$ superconducting flux qubits ($N=108$ in our case) subject to a transverse-field Ising model
\beqAlbash
H(t) = -A(t) \sum_{i\in V(G)} \sigma_{i}^x + B(t) H_{\mathrm{Ising}}\ ,
\label{eq:H}
\eeqAlbash
where now the $\sigma$'s are the standard Pauli spin-$1/2$ matrices. The annealing schedules are shown in Fig.~\ref{Albash-fig:DW1}. The system is cooled very nearly into the ground state of the transverse field at $t=0$, given that the operating temperature is much smaller than $A(0)$. The only programmable parameters are the total annealing time $t_a \in [5\mu\textrm{s},20\textrm{ms}]$, the local fields $h_i$, and the couplings $J_{ij}$, subject to the Chimera graph connectivity. All our experiments reported here were conducted with $t_a = 5\mu$s, all $h_i=0$, and all $J_{ij}$ chosen at random as $\pm 1$.

In the SSSV model qubits are replaced by planar rotors taking angle values $\theta_i\in [0, \pi]$. The Hamiltonian in Eq.~\eqref{eq:H} governs the dynamics of the system after the replacements $\sigma^x_i \mapsto \sin\theta_i$ and $\sigma^z_i \mapsto \cos\theta_i$:
\beqAlbash
H_{\textrm{SSSV}}(t) = -A(t) \sum_{i\in V(G)} \sin\theta_i + B(t) \left[\sum_{\{(i,j)\}\in E(G)} J_{ij} \cos\theta_i\cos\theta_j + \sum_{i \in V(G)} h_i \cos\theta_i\right] \ .
\label{eq:H_SSSV}
\eeqAlbash
 In Ref.~\cite{Smolin} Smolin \& Smith proposed to integrate the resulting Newton's equations of motion for the angles after the addition of a noise term, but it was shown in Refs.~\cite{q108,comment-SS} that this ``spin dynamics'' model correlates poorly with the DW1 success probability data. Rather than integrating the equations of motion Shin \textit{et al.} \cite{SSSV} proposed to thermalize all the angles at each time step. This is done according to Monte Carlo dynamics: an angle $\theta_i \in [0, \pi]$ is generated with uniform probability for the $i$-th rotor; if the new angle reduces the energy computed according to Eq.~\eqref{eq:H_SSSV} the update is accepted; if not, the update is accepted with probability $e^{- \beta \Delta E}$, where $\beta$ is the inverse temperature and $\Delta E$ is the energy change. A complete update of all the spins is called a sweep. Starting from $t=0$, after each sweep the time in classical SSSV Hamiltonian is incremented $t$ to $t+\delta t$, until reaching $t_a$.

SQA is a quantum Monte Carlo (QMC) algorithm, where the state at each QMC sweep is equilibrated before performing a small change in Hamiltonian, after which the state is again equilibrated and so forth.  In this manner Monte Carlo dynamics again governs the evolution of the system. In more detail, SQA performs a path-integral QMC simulation of a transverse field quantum Ising model. As we discuss in detail in Sec.~\ref{sec:SSSV-deriv}, the path-integral formulation maps the quantum spin system to a classical spin system by adding an extra spatial dimension of extent $\beta$. The QMC simulation then performs stochastic updates of this classical path-integral configuration. In SQA one updates the coupling parameters by following the same schedule as in QA. If this change happens slowly enough the QMC simulation equilibrates to the new couplings on a short timescale and always samples the canonical ensemble of the instantaneous Hamiltonian. The same is true for physical QA as intended to be embodied in the D-Wave devices. Thus, in the limit of sufficiently long annealing times, both SQA and QA sample from the same time-dependent ensemble. In the case of a rough energy landscape in a spin glass problem, both SQA and QA can be trapped in a local minimum. An avoided level crossing with small gaps has its origin in small tunneling matrix elements that in turn are due to large Hamming distances between two local minima. The large Hamming distance and small tunneling matrix elements mean that a tunneling event to reach the new local minimum is strongly suppressed also in the QMC simulation of SQA, hence making these instances hard for both SQA and QA.  For hard spin glass instances the simulation is then typically unable to explore the whole energy landscape and sampling is limited to the local thermal equilibrium around a local minimum of the free energy. 
While the hardness thus correlates between QA and SQA, it remains an open question whether a physical QA, operating at temperatures above that of the smallest gaps can be more efficient than SQA. Moreover, as we shall see here, a classical (mean-field) model such as SSSV can be an accurate approximation of SQA for the right ensemble of spin glass instances.

Differences among the individual superconducting flux qubits contribute to systematic errors in the DW1 results. To average out these errors we implemented different ``gauges", a technique introduced in Ref.~\cite{q-sig}. Starting from the original Hamiltonian, by replacing $J_{ij} \mapsto a_i a_j J_{ij}$ and $h_i \mapsto a_i h_i$, where each $a_i$ is chosen randomly as $+1$ or $-1$, we map the original Hamiltonian to a gauge-transformed Hamiltonian with the same energy spectrum but where the identity of each energy eigenstates is relabeled accordingly, as $\sigma_i^z \mapsto a_i \sigma_i^z$. When a gauge is programmed on the device, there is calibration noise in the $(h,J)$ values, i.e., the ideal $(h,J)$ values and the actual programmed $(h,J)$ value are not necessarily the same.  This calibration error remains fixed for the remaining number of runs at that gauge.  To model this error, we used Gaussian noise with mean 0 and standard deviation $\sigma$ on the $(h,J)$ values of the programmed qubits.  We used $16$ gauges of the DW1 data available with $1000$ runs per gauge and similarly for SSSV and SQA.  The SSSV and SQA models have two further parameters: the temperature $T$ in mK and the number of annealing sweeps $n$.  We investigated a wide range of values, and for brevity we label each simulation with its parameters as $(T,n,\sigma)$.  In the case where no noise is added, we denote this case by $\sigma = 0$.

\section{Energy-success probability distributions}
\label{sec:espd}

\begin{figure}[t]
	\subfigure[\ DW1]{\includegraphics[width=.32\columnwidth]{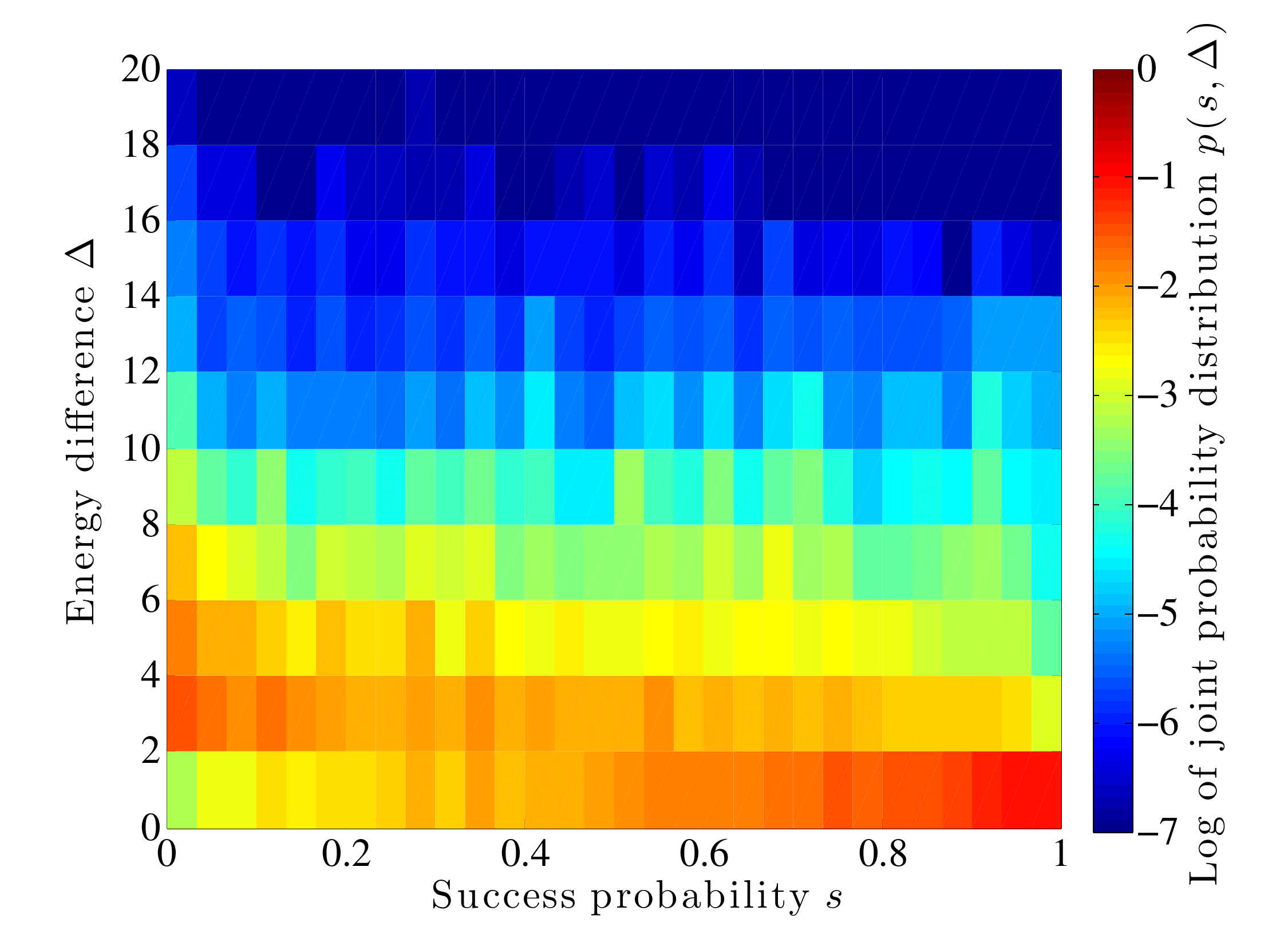}\label{Albash-fig:ES_DW1}}
	\subfigure[\ SSSV $(10.56, 150k, 0.05)$]{\includegraphics[width=.32\columnwidth]{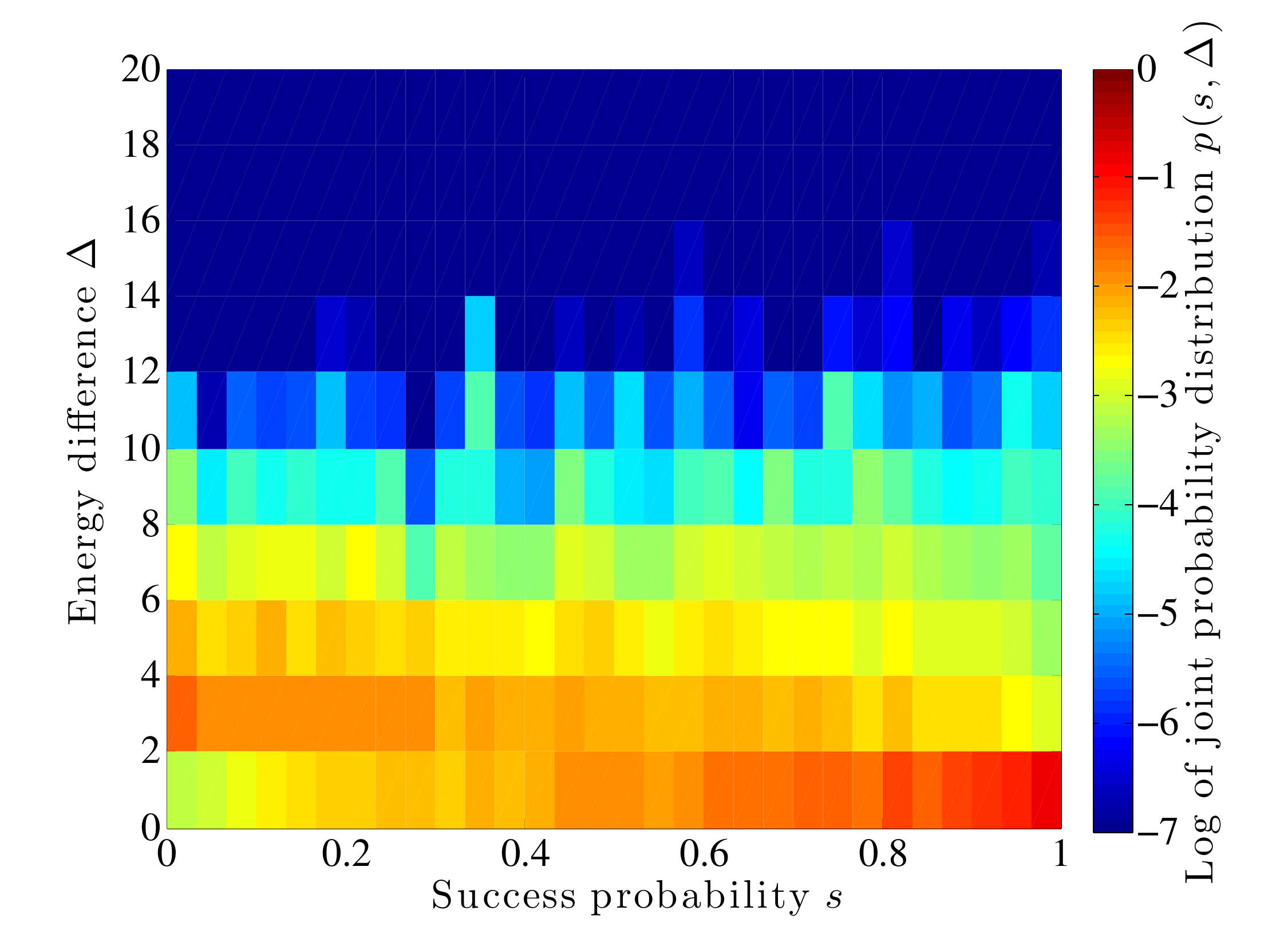}\label{Albash-fig:ES_SSSV}}
	\subfigure[\ SQA $(0.76, 10k, 0.05)$]{\includegraphics[width=.32\columnwidth]{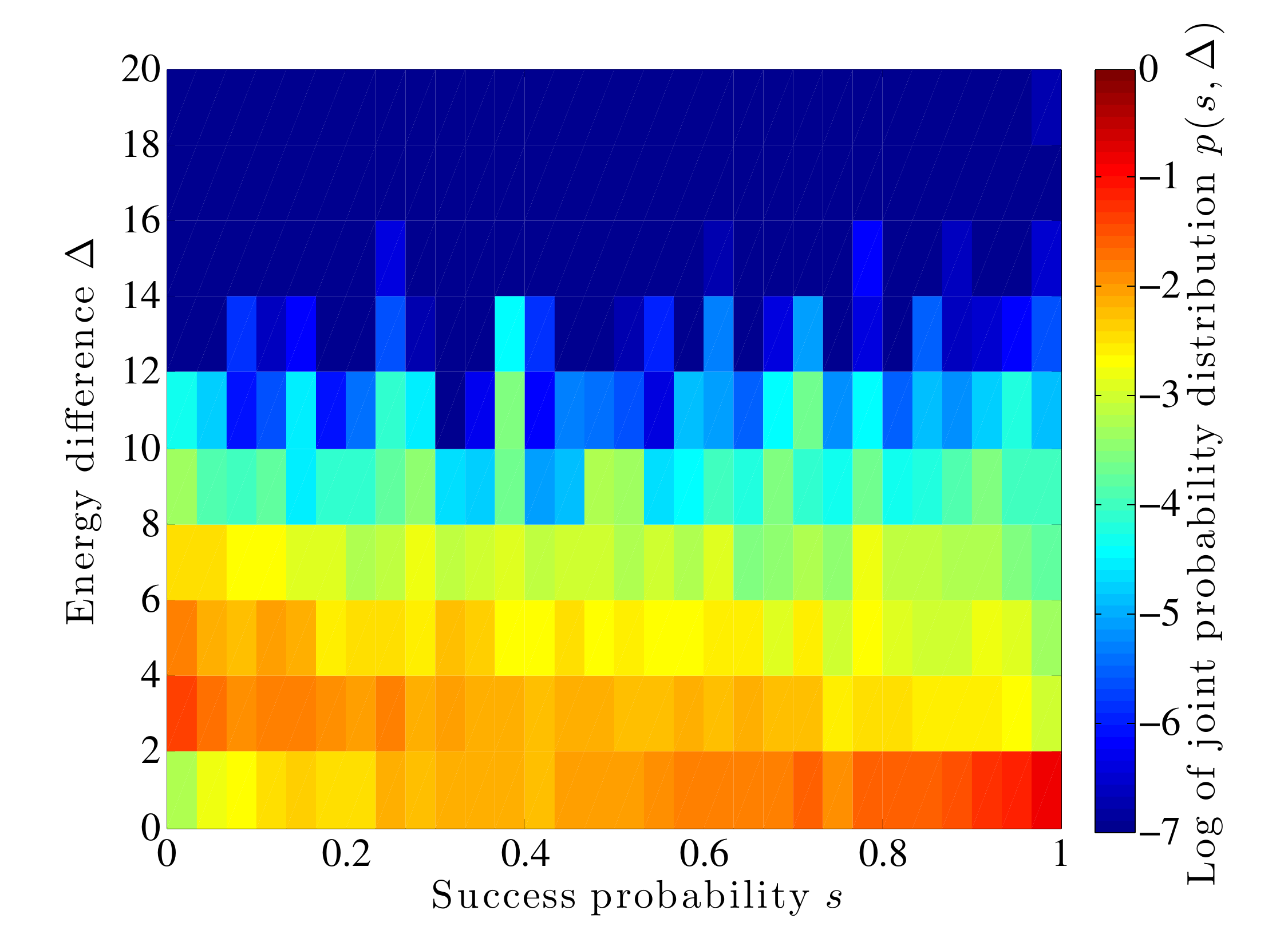}\label{Albash-fig:ES_SQA}}
	\caption{Joint energy-success probability distributions for (a) DW1, (b) SSSV, and (c) SQA. The success probability $s$ is the number of times the correct ground state was found for a given $108$-qubit random Ising instance in $1000$ runs. For all instances the Ising couplings $J_{ij}$ were chosen uniformly at random from $\pm 1$ and the local fields $h_i$ were set to zero. $\Delta$ is the energy difference from the ground state, such that $\Delta = 0$ corresponds to a ground state, while $\Delta > 0$ corresponds to an excited state. A total of $1000$ instances are shown; when the ground state was not found for a given instance, $\Delta$ represents the energy of the excited state that was found. The data is gauge-averaged over $16$ randomly chosen gauges (see text).  Panels (b) and (c) are labeled by their simulation parameters $(T,N,\sigma)$ corresponding to the temperature, number of Monte Carlo sweeps, and the standard deviation of the noise added to the Ising couplings and local fields (with zero mean). The three distributions are difficult to distinguish.}
	\label{Albash-fig:JPD}
\end{figure}

To go beyond Ref.~\cite{q108}, we do not restrict ourselves here to the ground state probabilities, but consider instead the entire output state distribution generated by the device and the models. Some such data was in fact already provided in Ref.~\cite{q108} (Supplementary Information), in terms of the joint distribution of success probabilities and energies of both the DW1 and SQA, and their similarity was another piece of evidence in favor of a quantum description of the DW1 (simulated annealing and spin dynamics disagreed with the DW1 data). We reproduce these results in Fig.~\ref{Albash-fig:JPD}, where we also include the SSSV model. For the chosen simulation parameters it is difficult to distinguish the experimental and simulations results, so we must conclude that this method is inappropriate for distinguishing the models and the experimental data. Note that we use the same set of previously \cite{q108} optimized SQA parameters throughout this work.

One reason that the joint energy-success probability distributions shown in Fig.~\ref{Albash-fig:JPD} do not distinguish the different methods is that when the data is presented in this manner it is not possible to make an instance-to-instance comparison.
Ideally, a comparison of which states each method finds would allow for a definitive measure of how correlated the different methods are.  However, the D-Wave processors suffer from random calibration errors that occur during the programming of the device \cite{Trevor}.  This results in a random deviation in the programmed Ising parameters from their ideal values.  Therefore, successive programming cycles effectively run a slightly different problem instance.  The final state and energy observed at the end of each run is highly sensitive to this effect and precludes a meaningful a state-to-state comparison.  However, the populations observed at a given energy level are more robust to this effect, and we next focus on this property.

\section{Comparing distributions via distance measures}
\label{sec:dist}
%

\subsection{Distance measure}
For each instance $i$ we define the probability distribution function $p_i(\Delta)$ of finding a state with energy difference $\Delta$ from the ground state energy $E_{0}$. This quantity can be computed separately for each method from our data: 
\beqAlbash
p_i(\Delta) = \frac{1}{N_E}\sum_{n=0}^{N_E} \delta_{E_{n} - E_{0},\Delta}\ ,
\eeqAlbash 
where $E_n$ is the energy of the $n$-th excited state and $N_E$ is the total number of excited energy levels observed for the given instance.
We then compute the total variation distance 
\begin{equation} 
\label{Albash-eqt:distance1}
\mathcal{D} \left(p,q \right) = \frac{1}{2} \sum_{x} \left| p(x) - q(x) \right|
\end{equation}
for a given instance $i$ between the probability distributions for DW1 and the different models, i.e., in Eq.~\eqref{Albash-eqt:distance1} we let $p = p_i^{\mathrm{DW1}}$ and $q=p_i^{\mathrm{SSSV}}$ or $q=p_i^{\mathrm{SQA}}$, and sum over $\Delta$.
To calculate this quantity and its associated error, we perform $b = 1000$ bootstraps (separately) on both the DW1 data and the model data (this mimics doing the experiment $1000$ times), calculate the distance between the $n$-th bootstrap for each, and then calculate the mean of the $b$ distances and their standard deviation.  This gives us an estimate of the distance and its error for the $i$-th instance.

 \begin{figure}[t]
 \centering
	\subfigure[]{\includegraphics[width=.34\columnwidth]{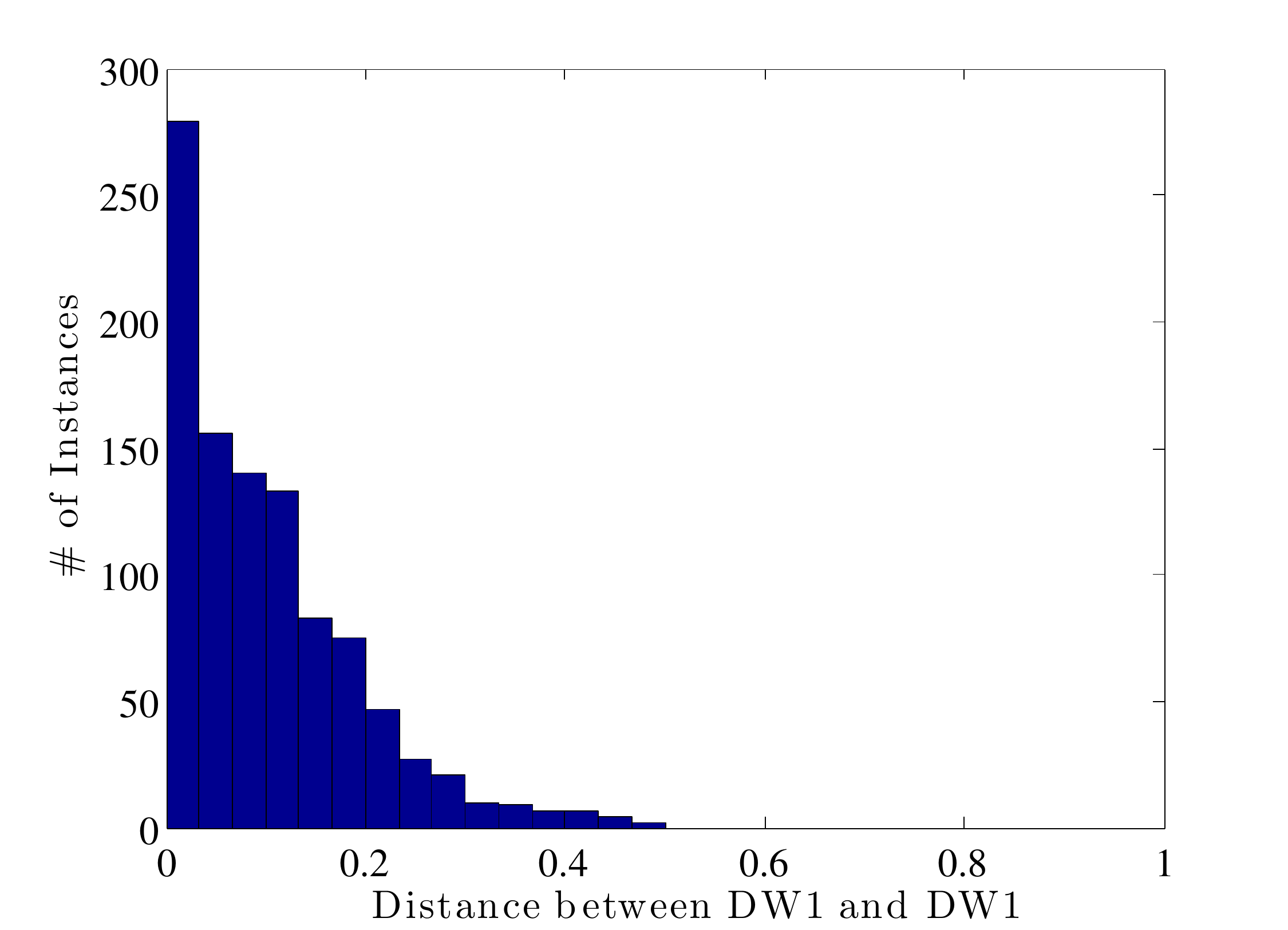} \label{Albash-fig:DW1-DW1}}
	\subfigure[\ SSSV: $(10.56, 150k, 0.05)$]{\includegraphics[width=.34\columnwidth]{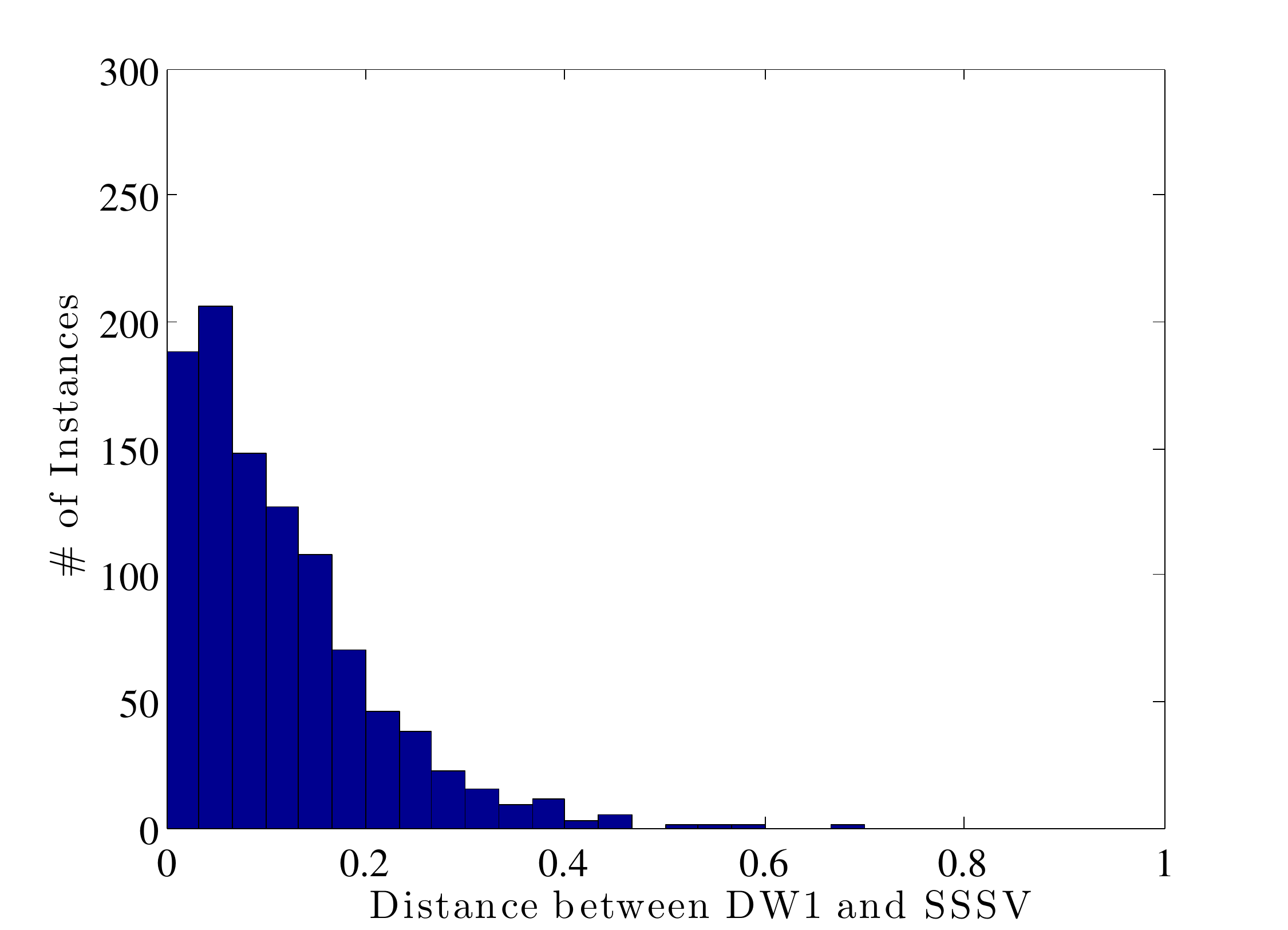}\label{Albash-fig:DW1-SSSV}}
	\subfigure[\ SQA: $(0.76, 10k, 0.05)$]{\includegraphics[width=.34\columnwidth]{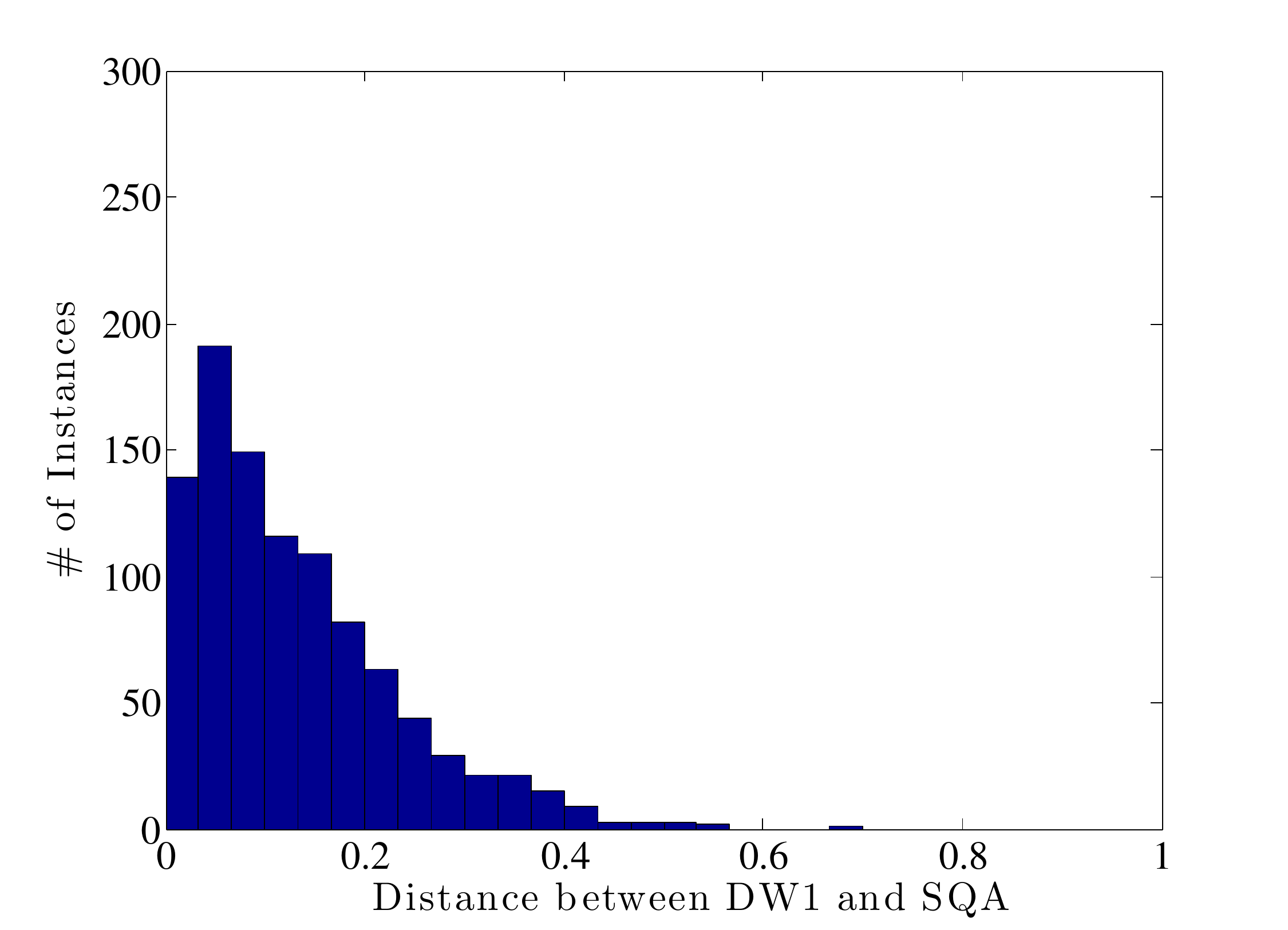}\label{Albash-fig:DW1-SQA}}
	\subfigure[]{\includegraphics[width=.34\columnwidth]{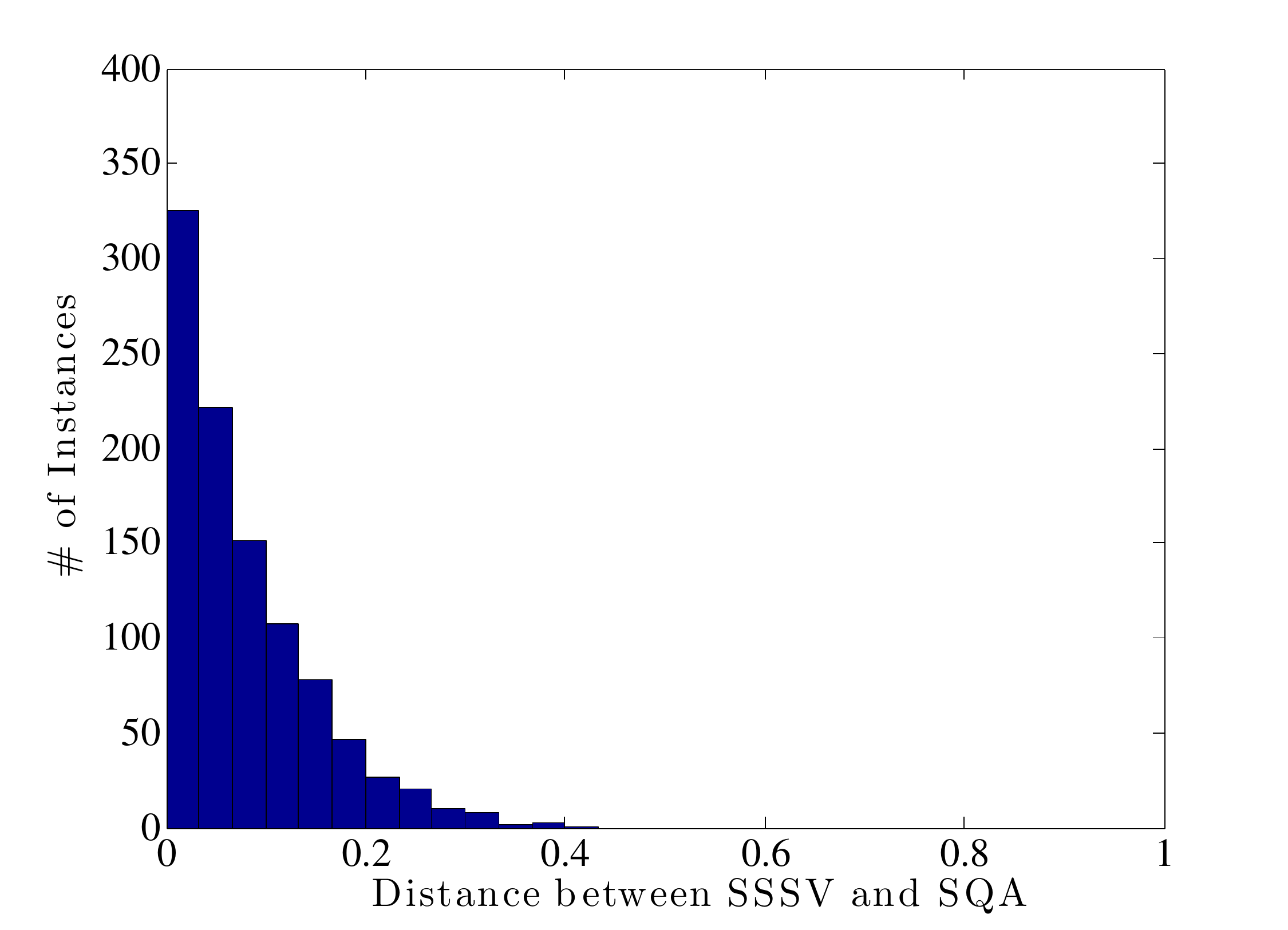}\label{Albash-fig:DistanceHistogram2}}
	\caption{Histograms of the total variation distances for the $1000$ random Ising instances.  (a) Distance between the DW1 and itself for two different sets of $8$ gauges. This measures how well the device correlates with itself.  (b) Distance between the DW1 and SSSV, with parameters $(10.56, 150k, 0.05)$.  (c) Distance between the DW1 and SQA, with parameters $(0.76, 10k, 0.05)$.  (b) and (c) use the median annealing schedules as depicted in Fig.~\ref{Albash-fig:DW1}(b). (d) Distance between SSSV and SQA for the $1000$ random Ising instances. Simulation parameters are $(10.56, 150k, 0.05)$ and $(0.76, 10k, 0.05)$, respectively, as in Fig.~\ref{Albash-fig:DistanceHistogram1}. Here and in all subsequent plots $16$ gauges are used except when comparing the DW1 with itself, for which we always used two different sets of $8$ gauges.}
	\label{Albash-fig:DistanceHistogram1}
\end{figure} 
%

\subsection{Correlation using the total variation distance}

In order to determine the reliability of this distance measure for our data set, we checked how well the DW1 correlates with itself by calculating the distance between two different sets of 8 gauges.  As shown in Fig.~\ref{Albash-fig:DW1-DW1}, we find that the histogram of distances is peaked around the first bin centered at $1/60$ (the bin size $1/30$ is determined by $\sim 1/\sqrt{1000}$, where $1000$ is the number of instances used).  There remains a substantial tail, which we attribute to the inherent noisiness of the device.  Nevertheless, we can use this as a basis of how correlated the different models are with the device, with any model matching this behavior being as correlated with the device as the device is with itself.

\begin{figure}[t]
\centering
	\subfigure[\ $(10.56, 150k, 0.075)$]{\includegraphics[width=.32\columnwidth]{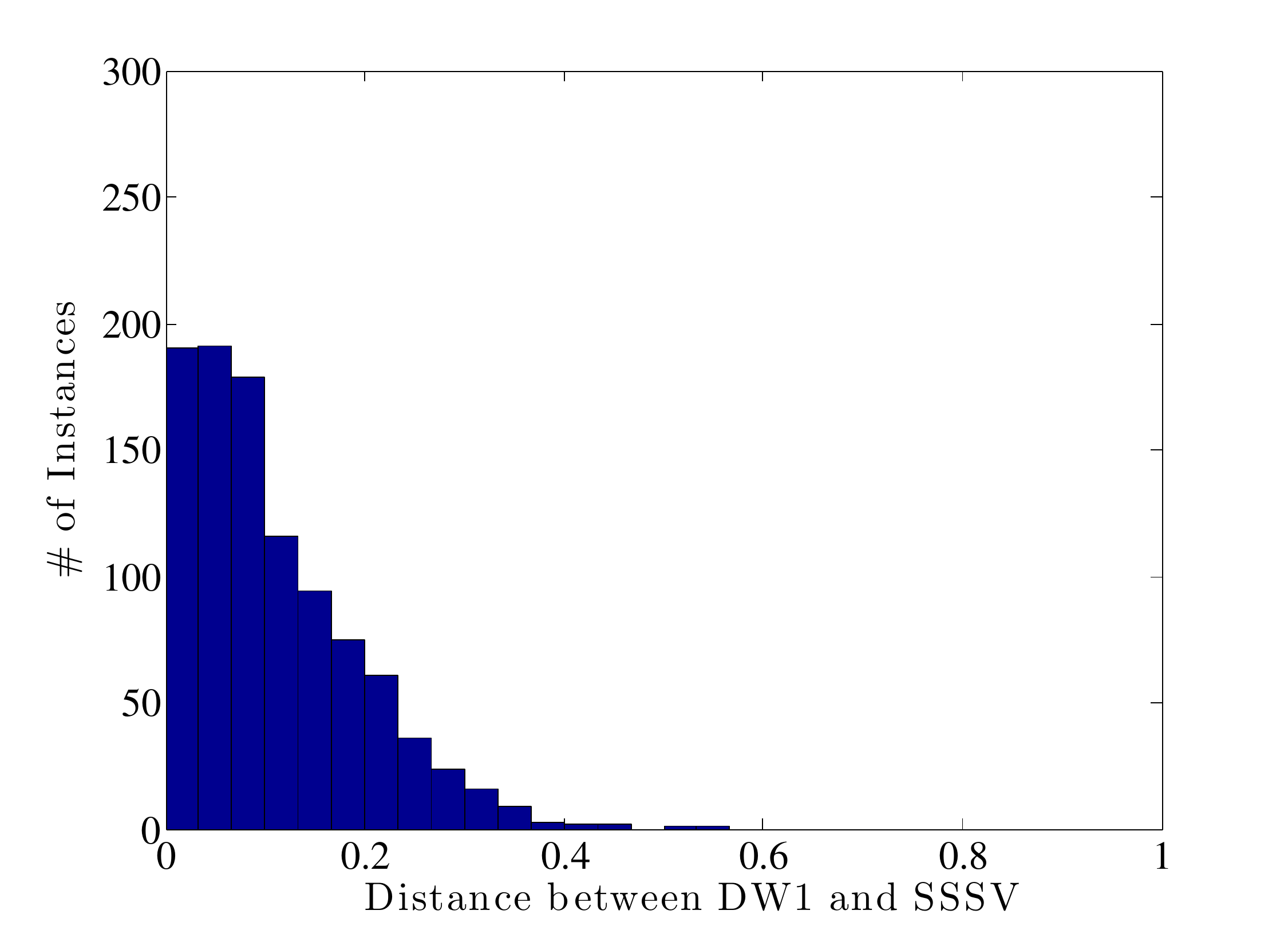} \label{Albash-fig:DW1-SSSV-075}}
	\subfigure[\ $(10.56, 100k, 0.05)$]{\includegraphics[width=.32\columnwidth]{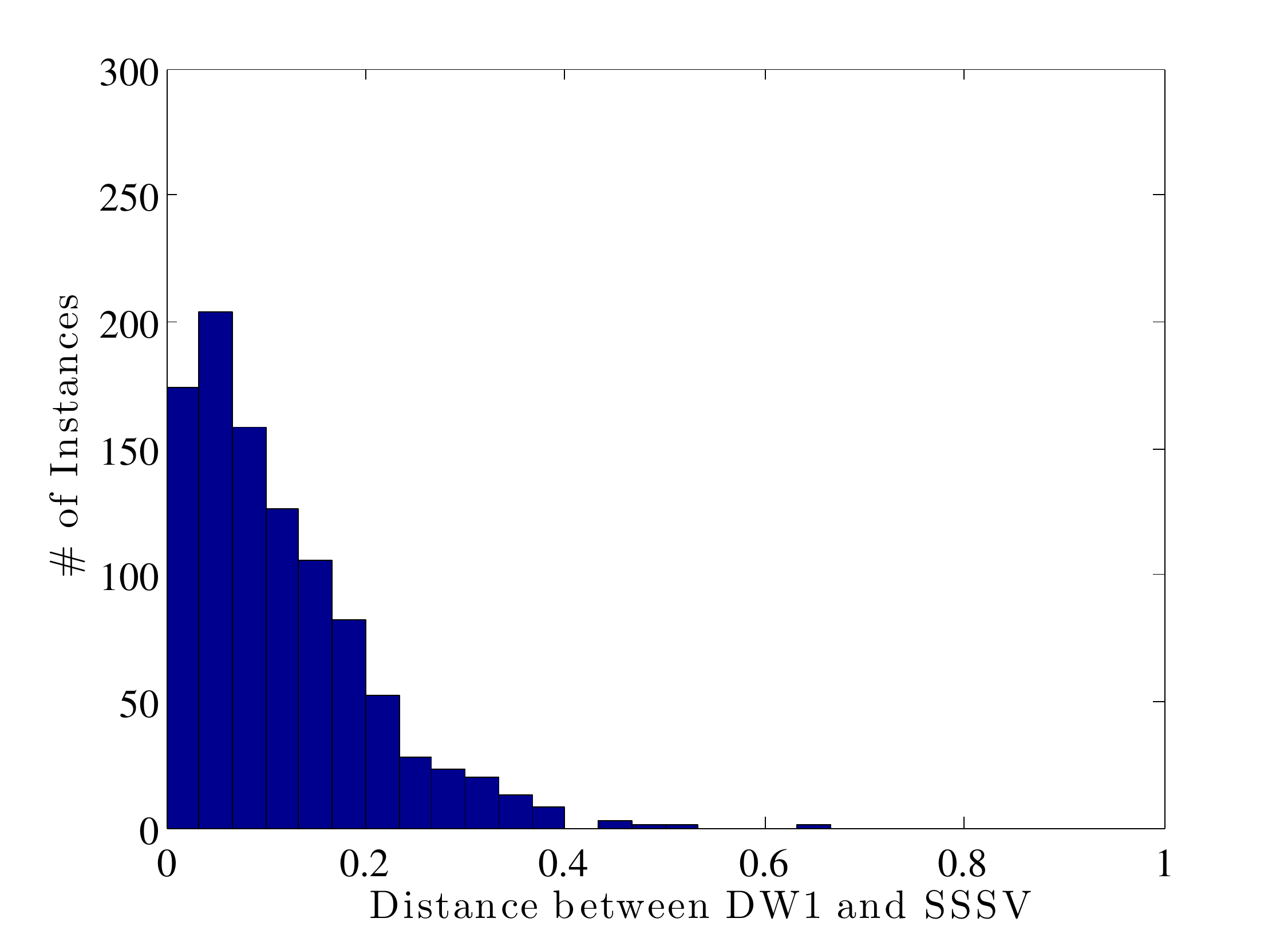}\label{Albash-fig:DW1-SSSV-150}}
	\subfigure[\ $(10.56, 200k, 0.05)$]{\includegraphics[width=.32\columnwidth]{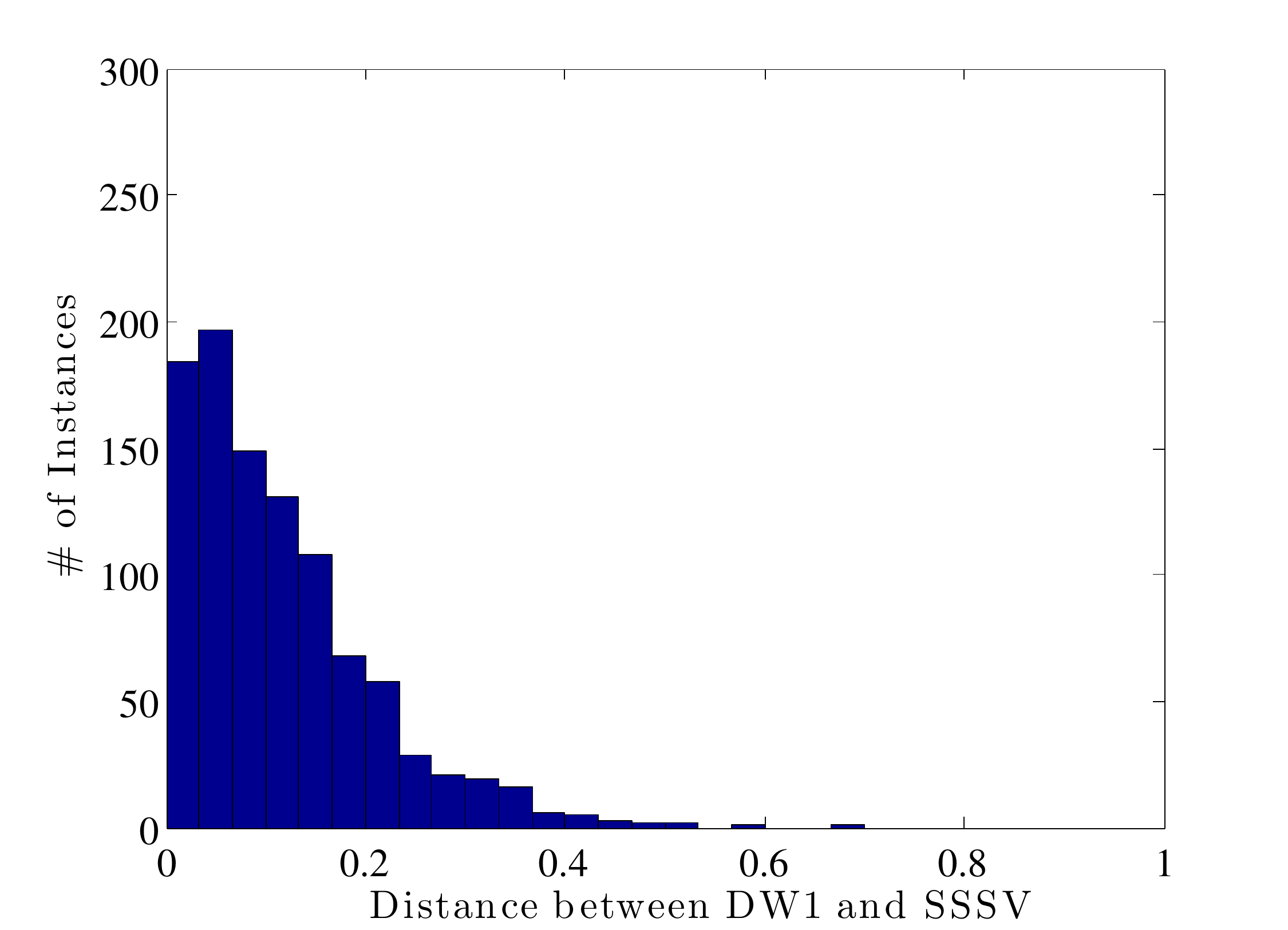}\label{Albash-fig:DW1-SSSV-200}}
	\subfigure[\ $(10.56, 150k, 0.05)$, $A_i$]{\includegraphics[width=.32\columnwidth]{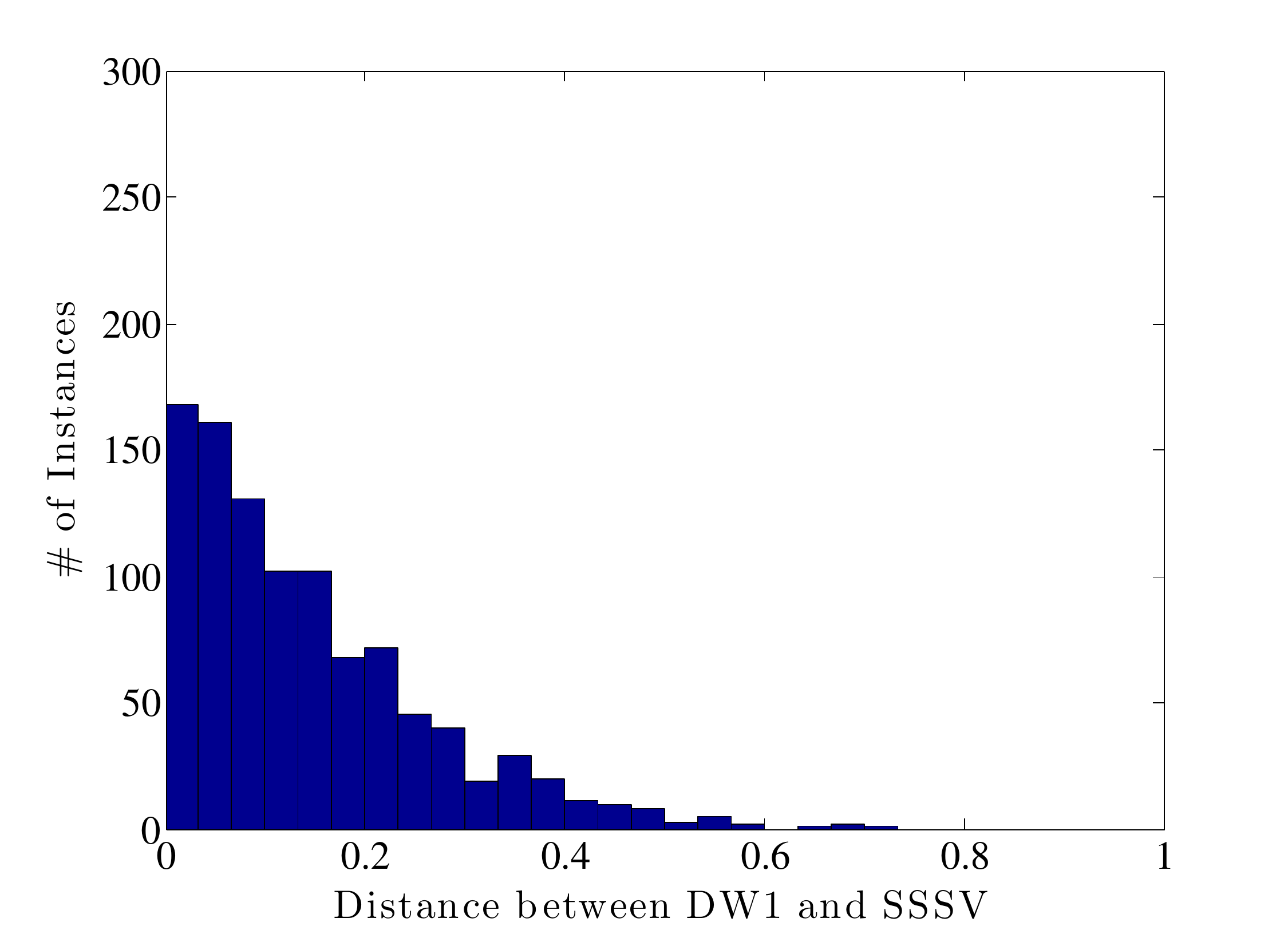} \label{Albash-fig:DW1-SSSV-TQubit}}
	\subfigure[\ $(10.56, 150k, 0.05)$, $\chi$]{\includegraphics[width=.32\columnwidth]{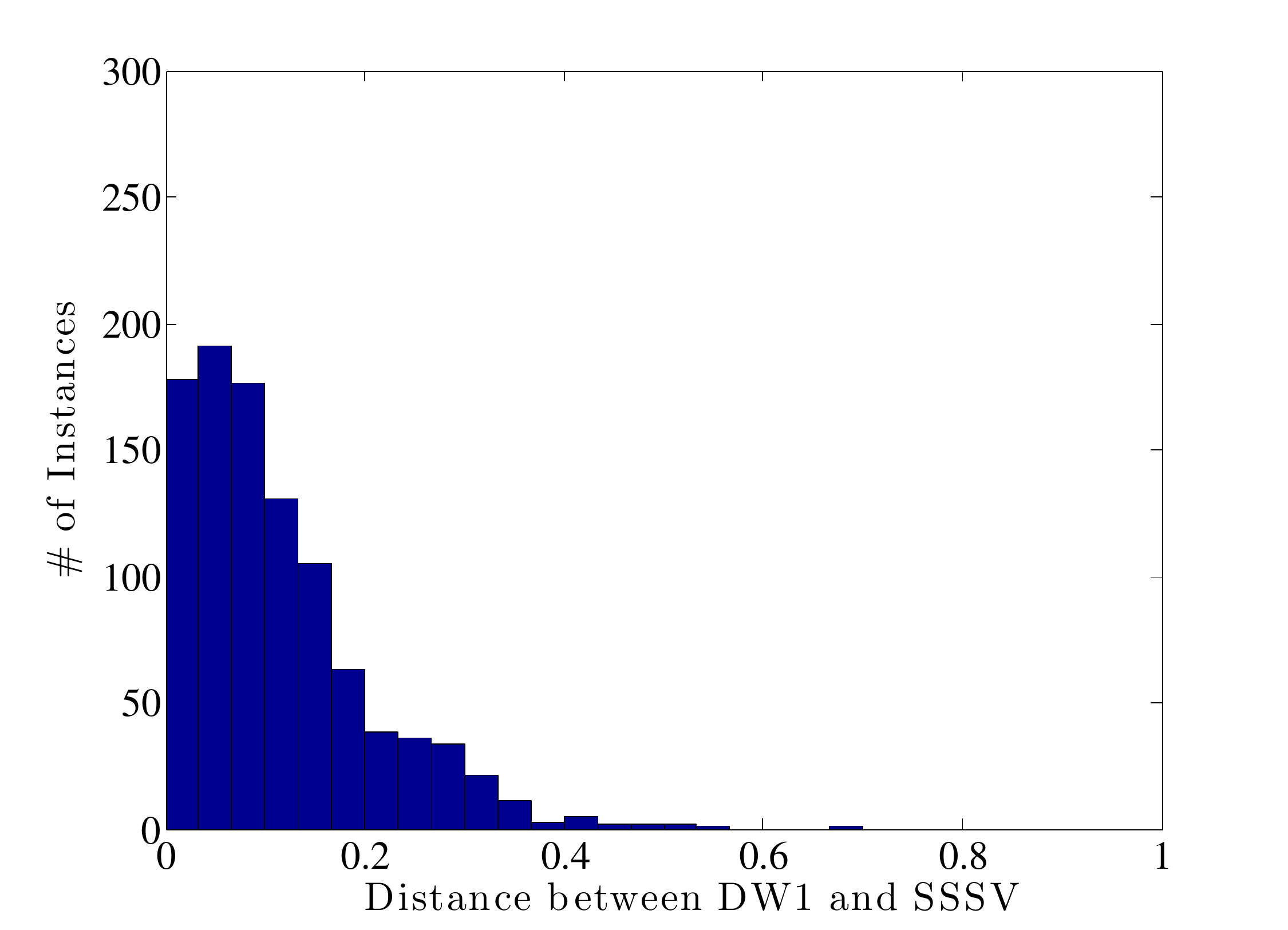}\label{Albash-fig:DW1-SSSV-Cross}}
	\subfigure[\ SQA: $(0.76, 10k, 0)$]{\includegraphics[width=.32\columnwidth]{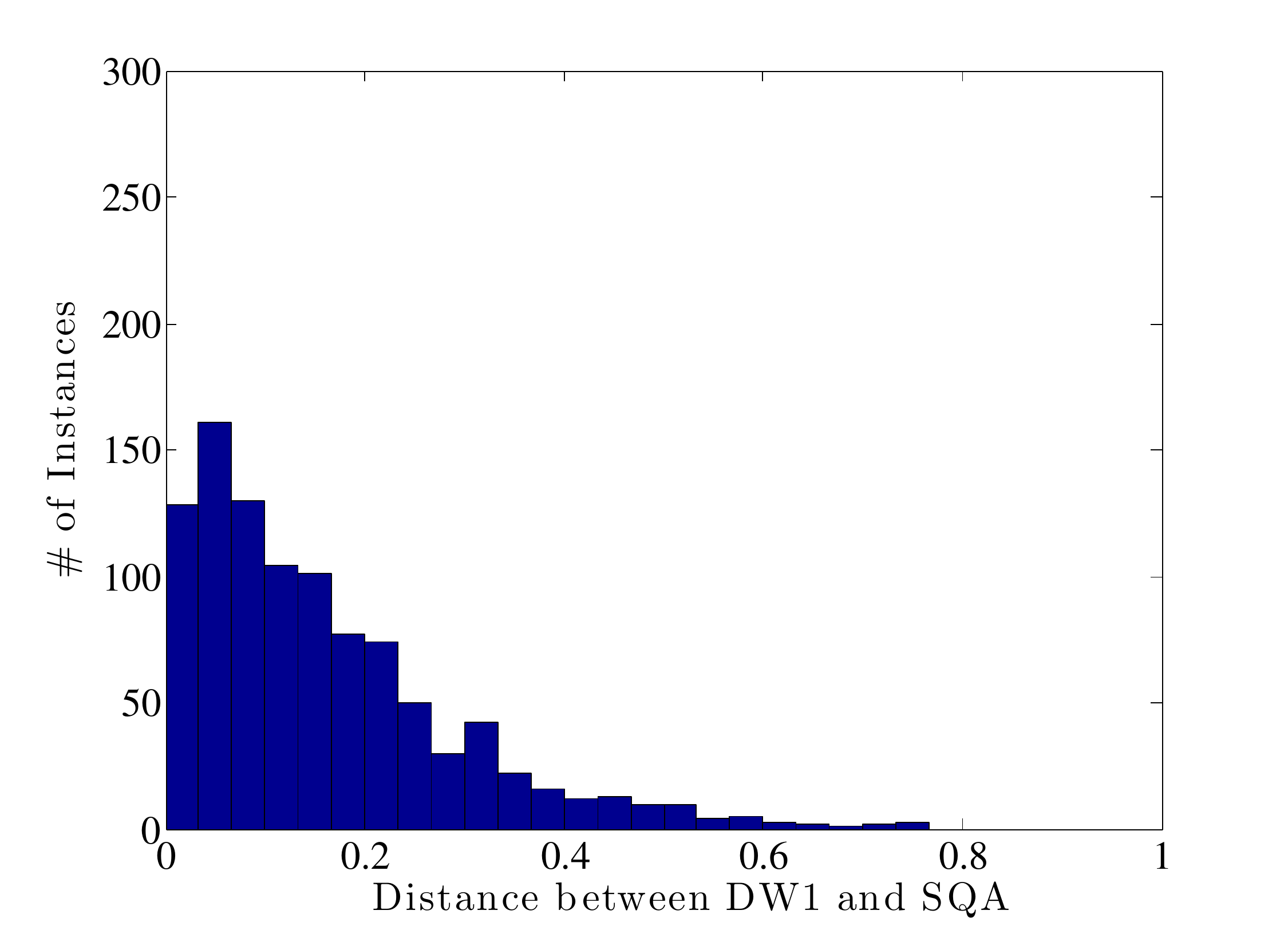}\label{Albash-fig:DW1-SQA_NoNoise}}
	\caption{Histograms of the total variation distances for the $1000$ random Ising instances for different simulations parameters.  (a) Larger standard deviation for the Gaussian noise on the local fields and couplings.  (b) and (c) Larger number of sweeps.  (d) With individual transverse field annealing schedules $A_i$, as shown in Fig.~\ref{Albash-fig:DW1}(c).  (e)  With crosstalk $|\chi| =0.05$ between qubits as modeled in Ref.~\cite{q-sig}. Panel (f) shows the SQA result without noise; the distance is larger than in the comparable case with noise, shown in Fig.~\ref{Albash-fig:DW1-SQA}. $16$ gauges were used for each method.}
	\label{Albash-fig:DistanceHistogram3}
\end{figure}   
\begin{figure}[t]
\centering
	\subfigure[\ median \textit{vs } individual $A_i$]{\includegraphics[width=.32\columnwidth]{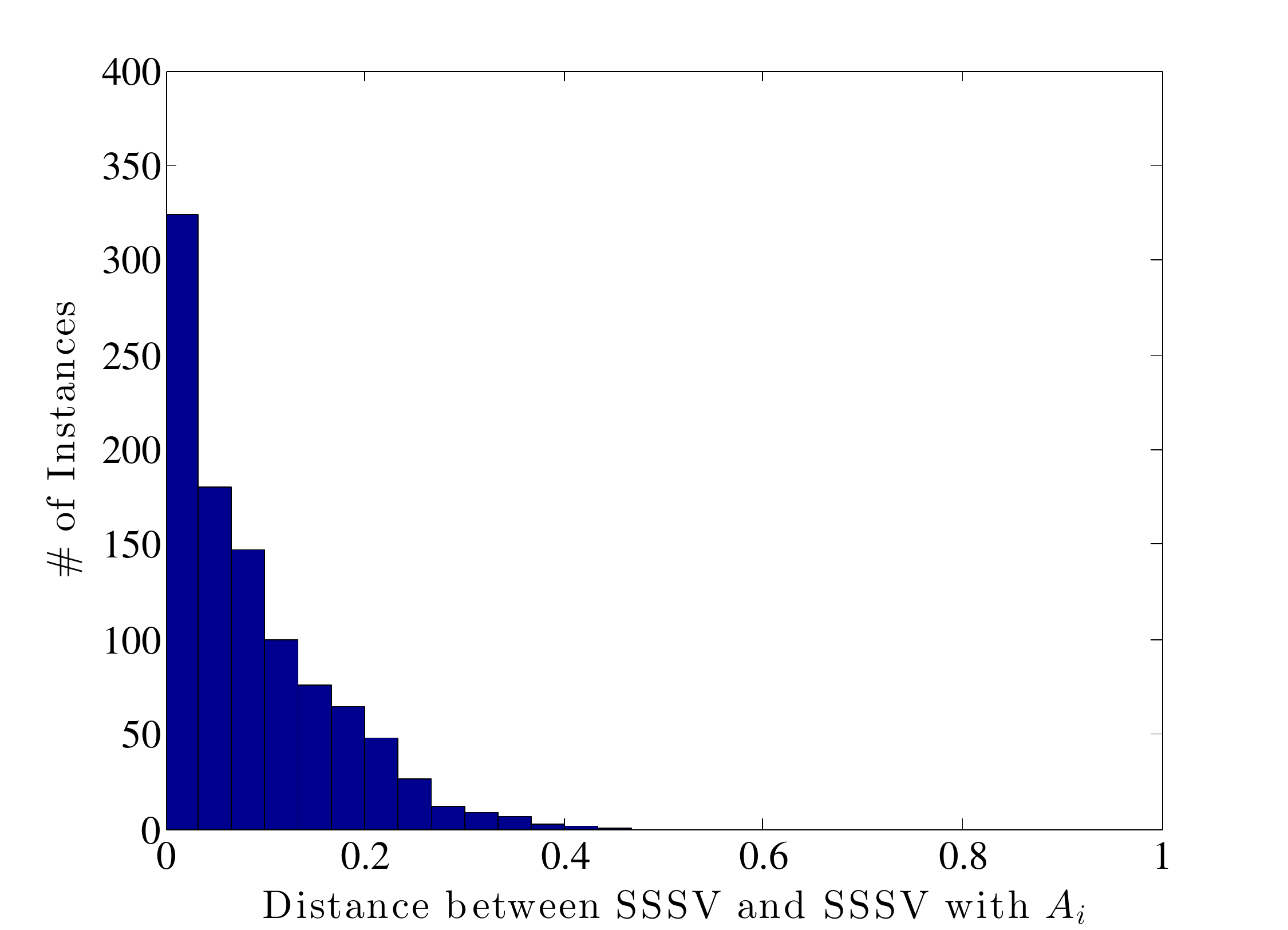} \label{Albash-fig:DW1-SSSV-075}}
	\subfigure[\ with \textit{vs } without crosstalk]{\includegraphics[width=.32\columnwidth]{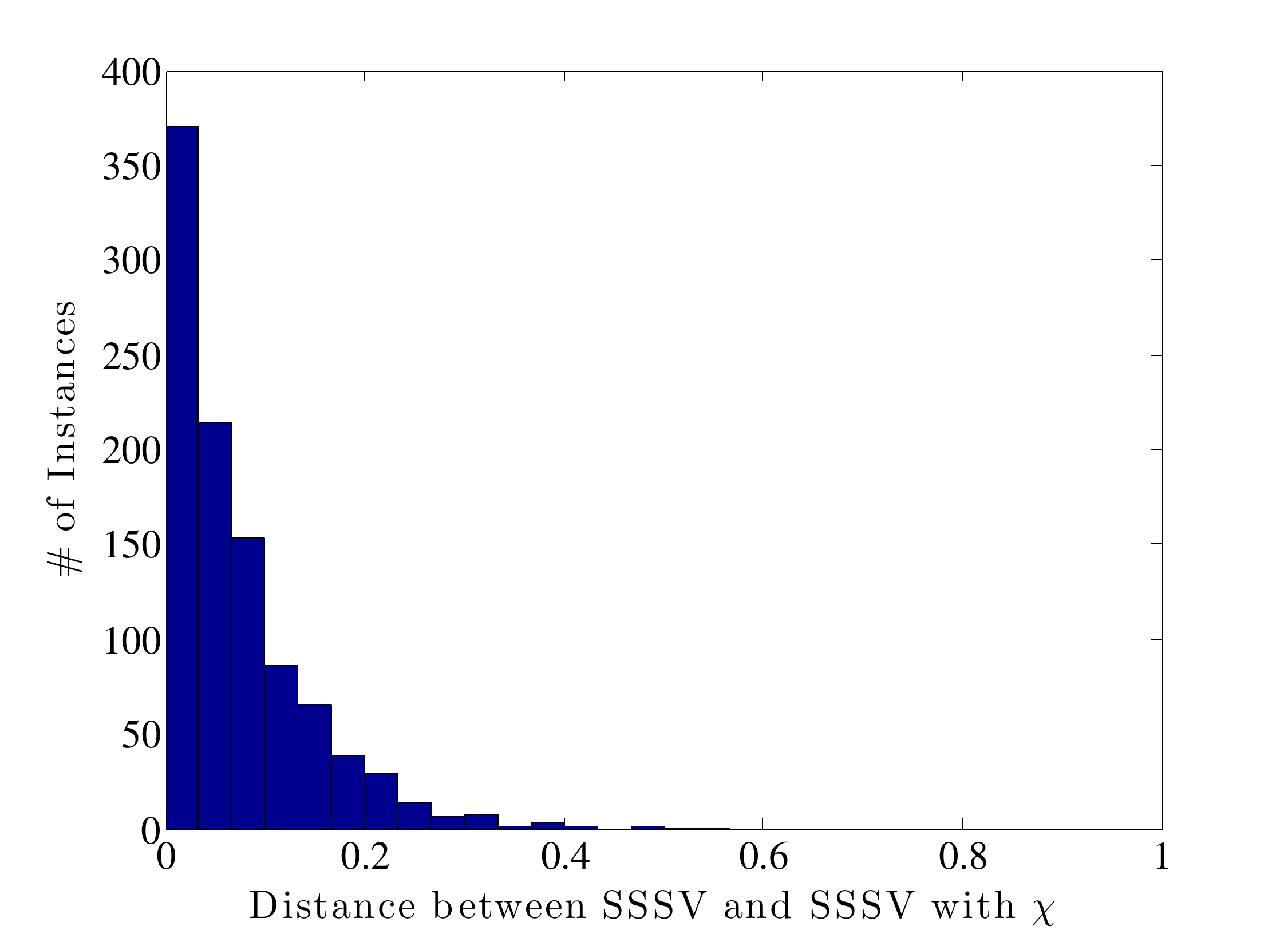}\label{Albash-fig:SSSV-SSSV-chi}}
	\caption{Histograms of the SSSV total variation distances for the $1000$ random Ising instances for different simulations parameters.  (a) SSSV with median annealing schedules [Fig.~\ref{Albash-fig:DW1}(b)] against SSSV with individual transverse field annealing schedule [Fig.~\ref{Albash-fig:DW1}(c)].  (b) SSSV with no crosstalk against SSSV with crosstalk between qubits as modeled in Ref.~\cite{q-sig} with $|\chi| = 0.05$. $16$ gauges were used for each method.}
	\label{Albash-fig:DistanceHistogram4}
\end{figure}   
Figures~\ref{Albash-fig:DW1-SSSV} and ~\ref{Albash-fig:DW1-SQA} are the total variation distance results for SSSV and SQA \textit{vs } DW1, respectively, where we have numerically optimized the values for each model's parameters so as to minimize this distance measure (see below).  Neither the SSSV nor the SQA histogram of distances has a strong peak at the smallest bin, indicating that while SSSV and SQA correlate well with the full state statistics of the DW1, the correlation is not perfect and there are statistically significant differences. However, SSSV and SQA correlate very strongly with each other, as shown in Fig.~\ref{Albash-fig:DistanceHistogram2}, indicating that their correlation extends to the excited state spectrum as well.

The parameter values used in Figs.~\ref{Albash-fig:DW1-DW1}-\ref{Albash-fig:DW1-SQA} were optimized in the sense that modifying the standard deviation of the Gaussian noise on the local fields and couplings, or modifying the number of sweeps, only reduced the correlation with the DW1 data. This is illustrated in Fig.~\ref{Albash-fig:DistanceHistogram3}. In addition, we have tried additional noise sources to test for possible improvements to the correlation.  We used individual transverse field annealing schedules and included crosstalk between qubits, neither of which significantly enhanced the correlation with the DW1 data (see Fig.~\ref{Albash-fig:DistanceHistogram4}).  In fact, the data with these additional noise sources correlates well with the data without these noise sources, suggesting that their effect is small.

We conclude that neither SQA nor SSSV correlate well with the DW1 when tested over the entire excited state spectrum, yet they correlate strongly with each other. This conclusion is robust to parameter variations.

\section{Ground state distributions and correlations}
\label{sec:GS}
%
 \begin{figure}[b]
 \centering
 	\subfigure[]{\includegraphics[width=.32\columnwidth]{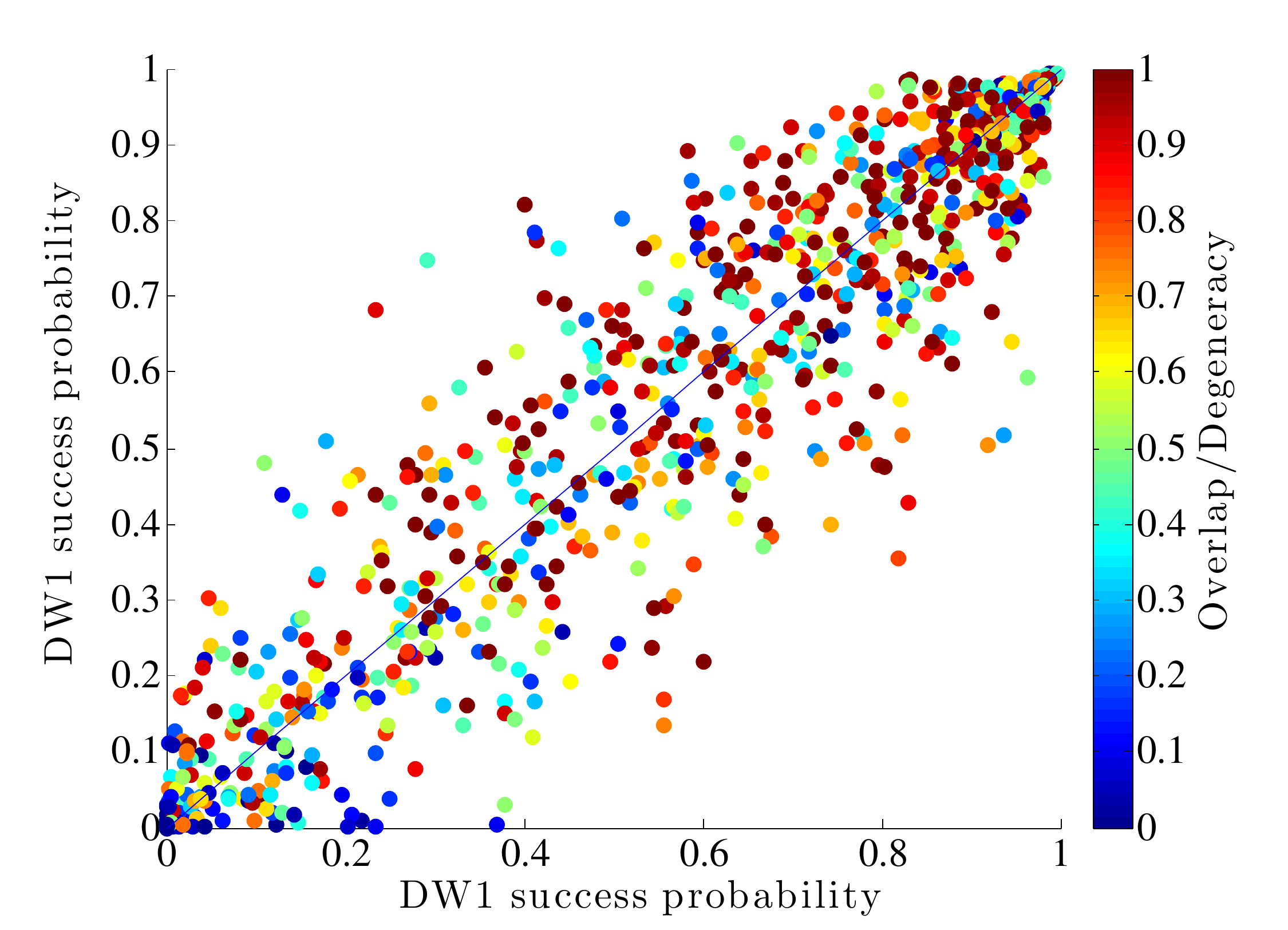}\label{Albash-fig:Overlap-DW1-DW1}}
	\subfigure[]{\includegraphics[width=.32\columnwidth]{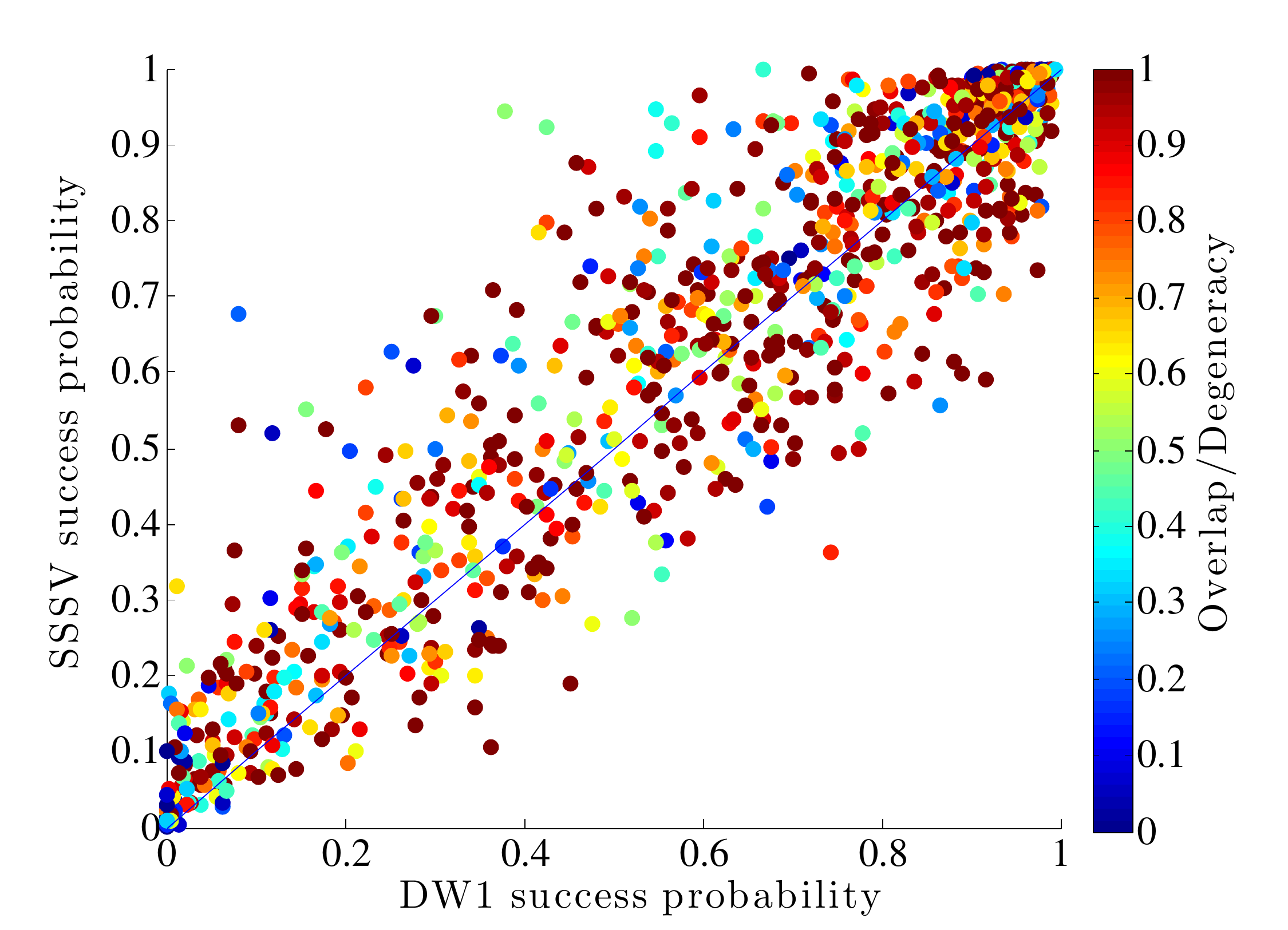}\label{Albash-fig:Overlap-DW1-SSSV}}
	\subfigure[]{\includegraphics[width=.32\columnwidth]{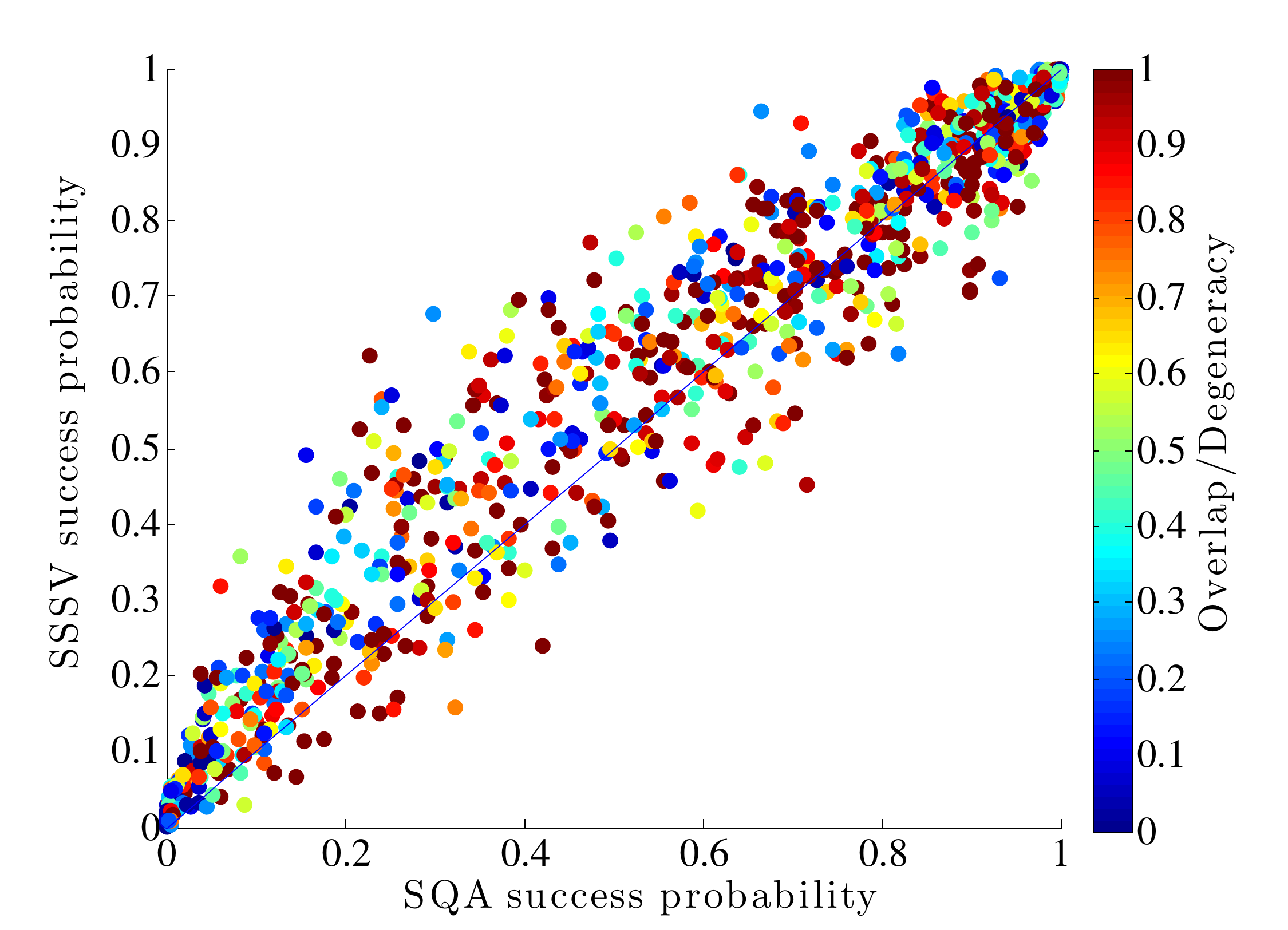}\label{Albash-fig:Overlap-SSSV-SQA}}
	\caption{Shown are the correlations between success probabilities of (a) DW1 \textit{vs } DW1, (b) SSSV \textit{vs } DW1, (c) SQA \textit{vs } SSSV. The SSSV and SQA parameters are optimized as in Fig.~\ref{Albash-fig:DistanceHistogram1}: $(10.56,150k,0.05)$ and $(0.76,10k,0.05)$.}
	\label{Albash-fig:Scatter0}
\end{figure}
 \begin{figure}[t]
 \centering
	\subfigure[\ DW1 \textit{vs } DW1]{\includegraphics[width=.32\columnwidth]{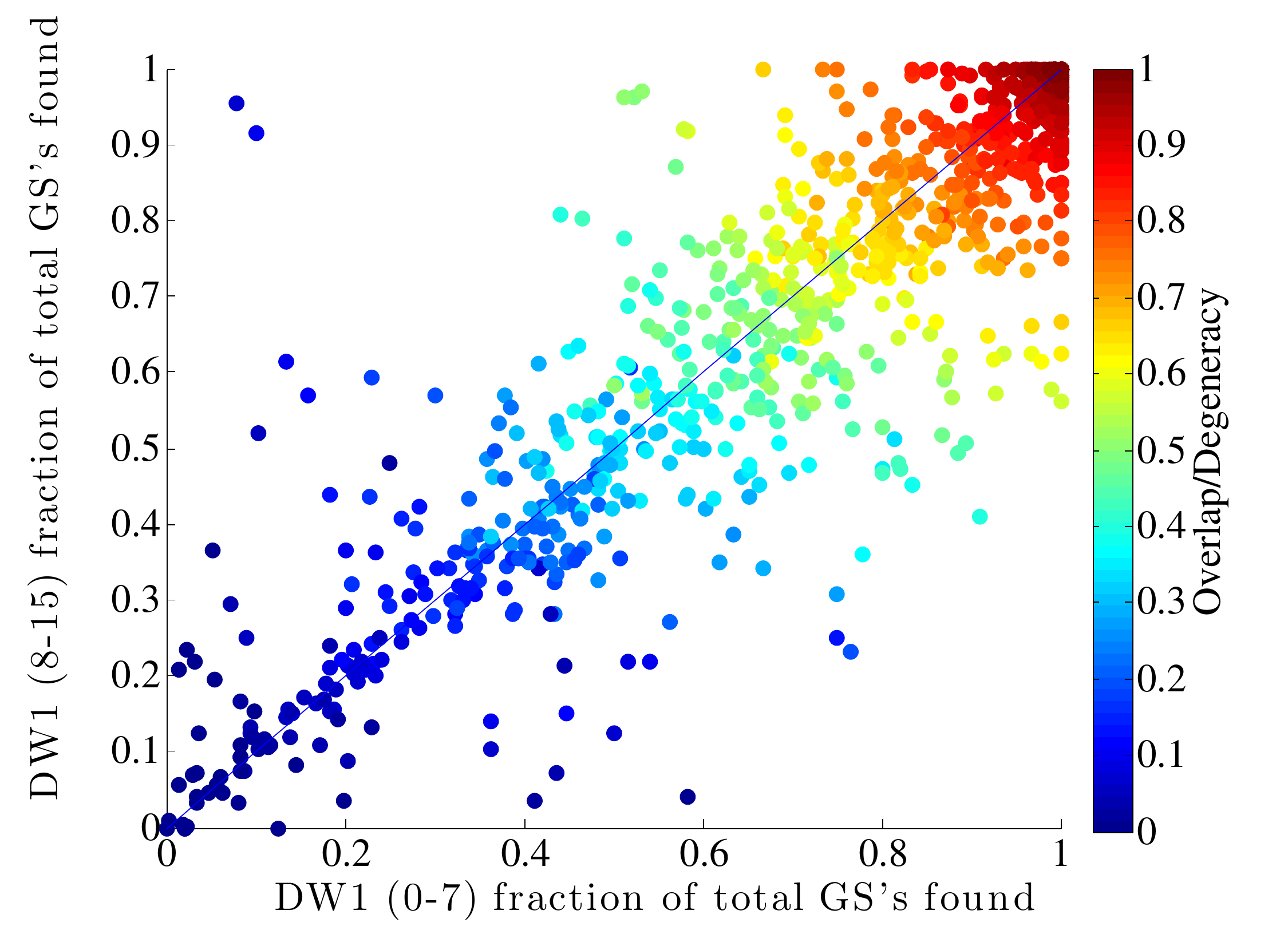}\label{Albash-fig:Fraction-DW1-DW1}}
	\subfigure[\ SSSV: $(10.56,150k,0.05)$]{\includegraphics[width=.32\columnwidth]{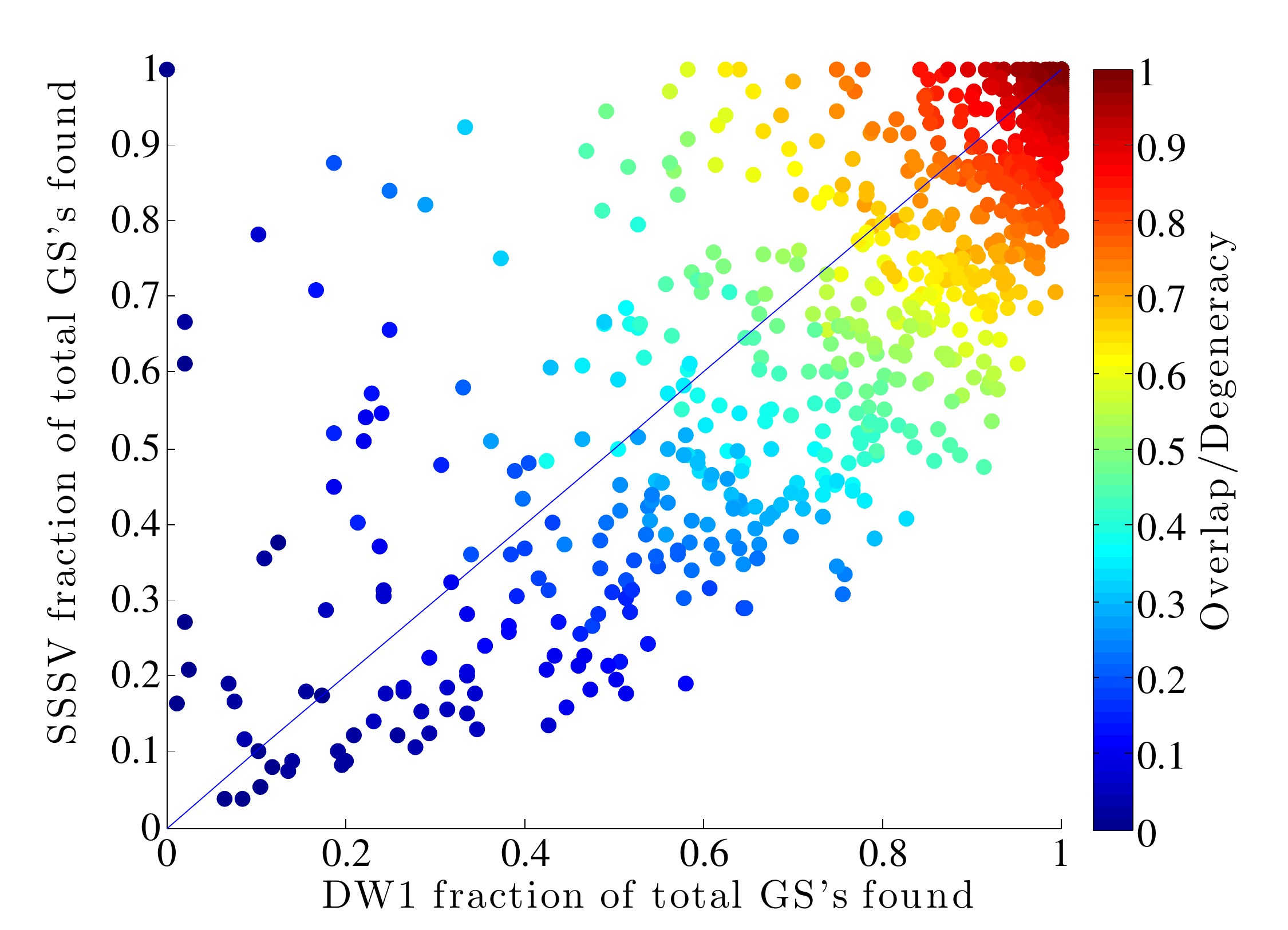}\label{Albash-fig:Fraction-DW1-SSSV}}
	\subfigure[\ SQA: $(0.76,10k,0.05)$]{\includegraphics[width=.32\columnwidth]{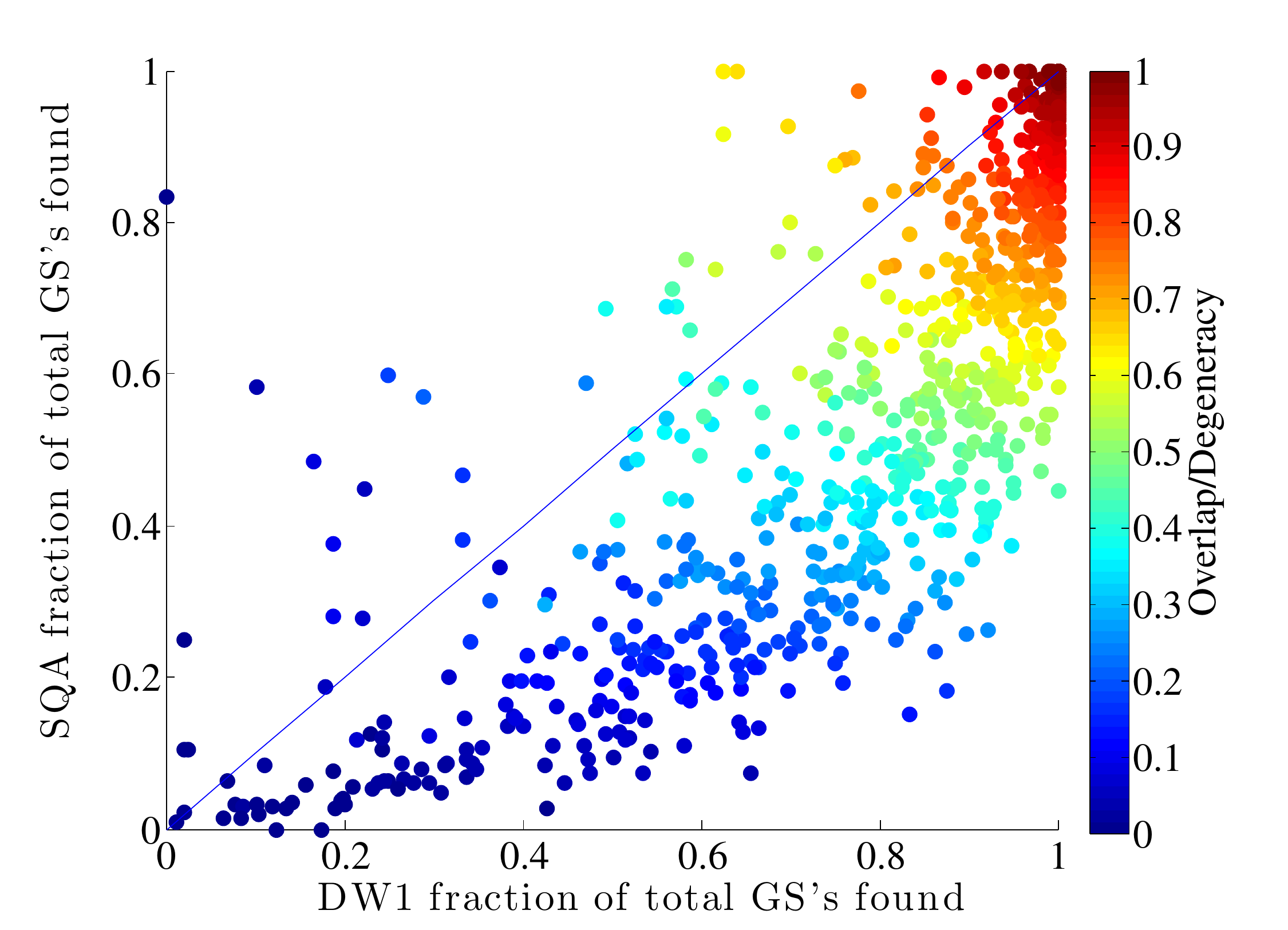}\label{Albash-fig:Fraction-DW1-SQA-beta=10}}
		\subfigure[\ SQA: $(2.54,10k,0.05)$]{\includegraphics[width=.32\columnwidth]{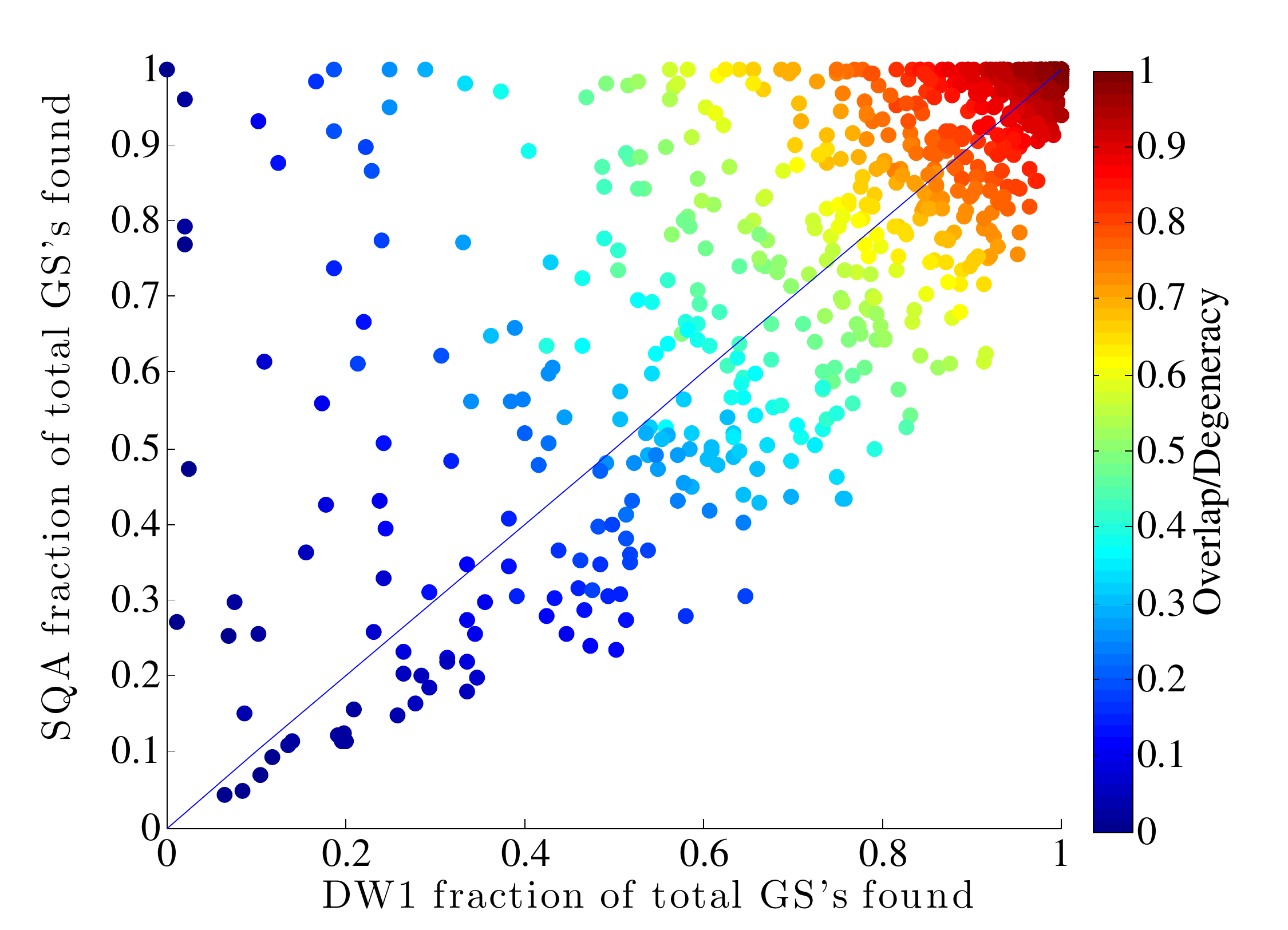}\label{Albash-fig:Fraction-DW1-SQA-beta=3}}
	\subfigure[\ SSSV: $(10.56,150k,0.05)$, SQA: $(0.76,10k,0.05)$]{\includegraphics[width=.32\columnwidth]{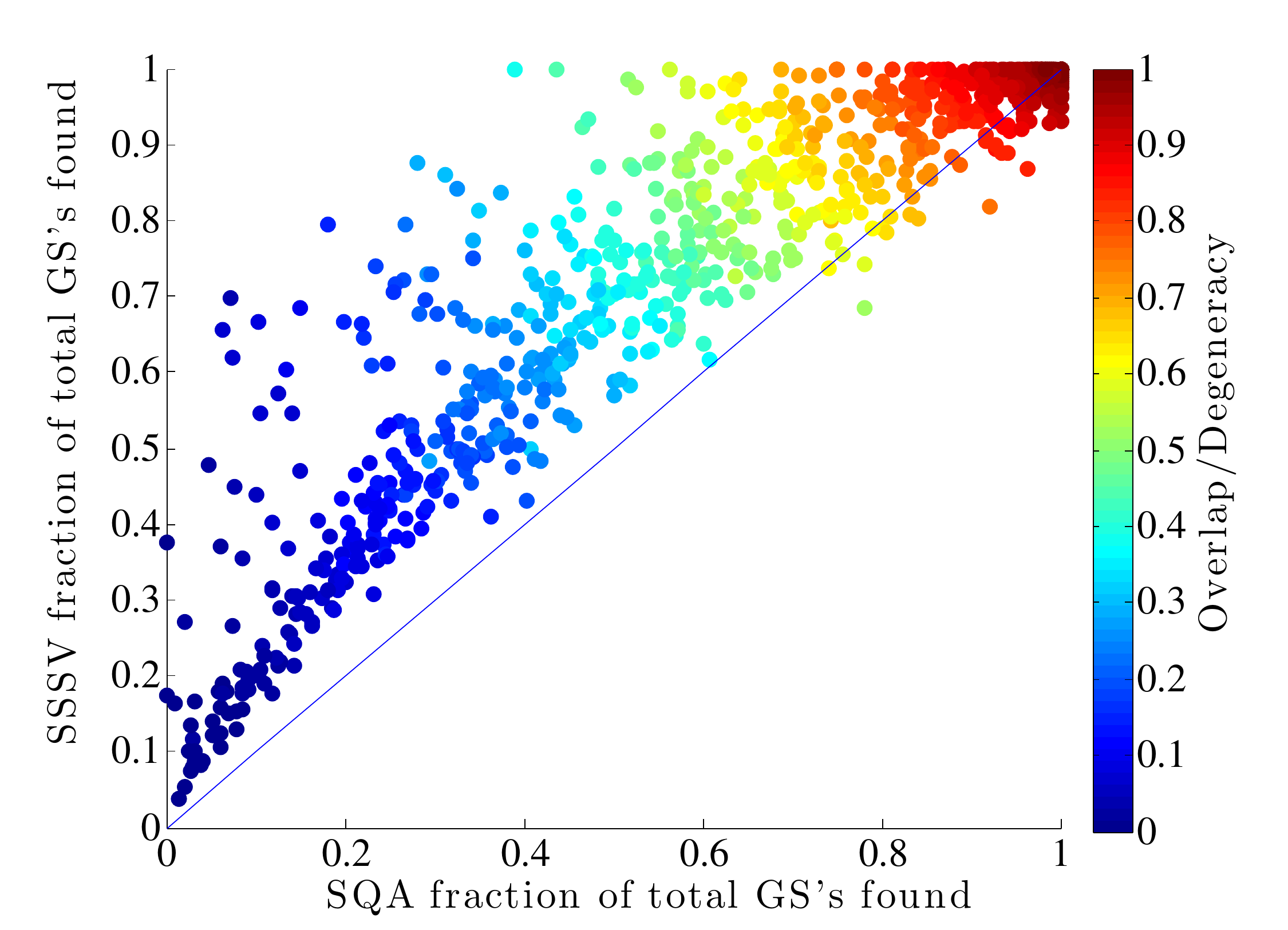}\label{Albash-fig:Fraction-SSSV-SQA}}
	\subfigure[\ SSSV: $(10.56,150k,0.05)$, SQA: $(2.54,10k,0.05)$]{\includegraphics[width=.32\columnwidth]{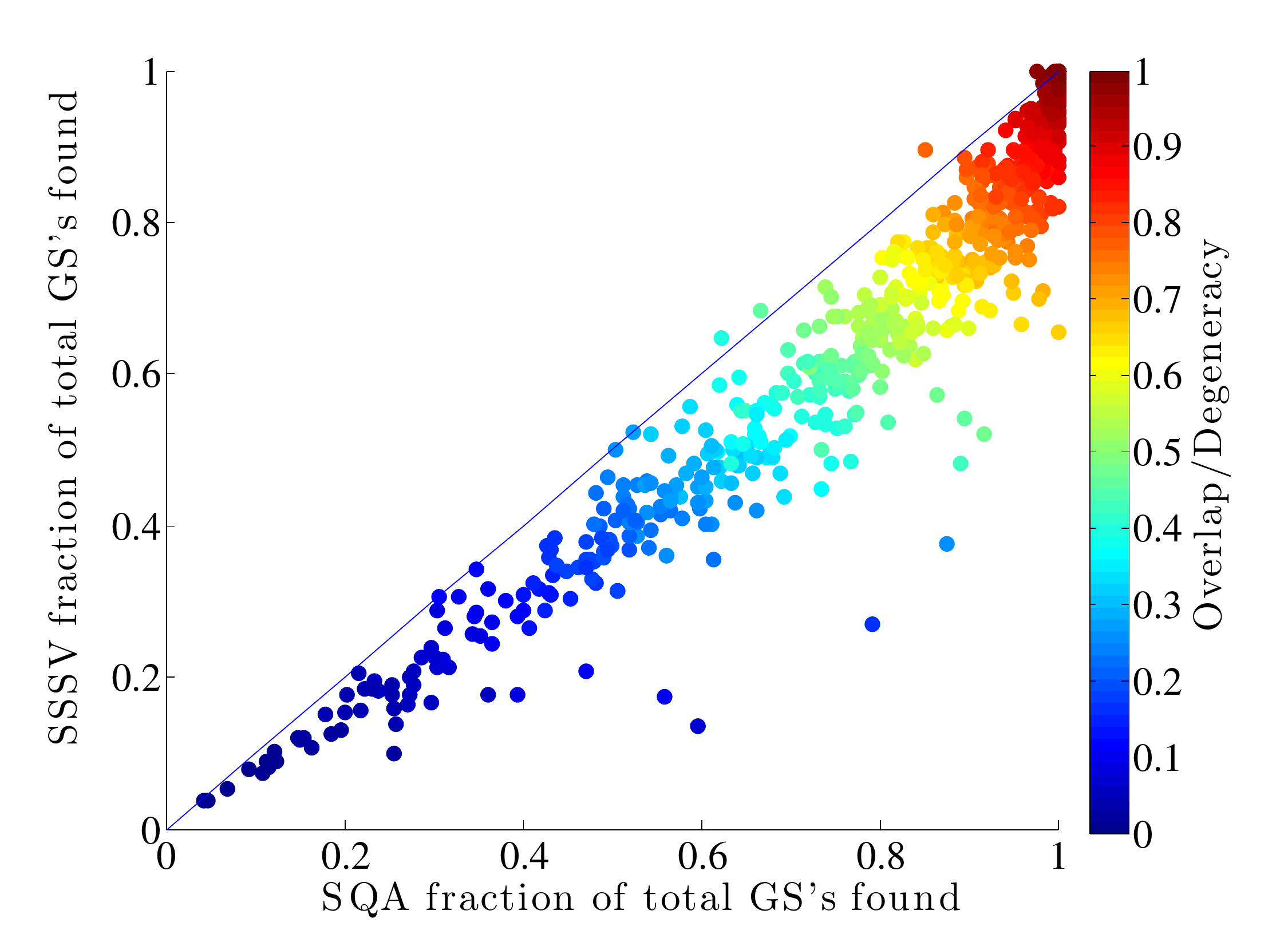}\label{Albash-fig:Fraction-SSSV-SQA-beta=3}}
	\caption{Shown are the correlations between the fraction of total number of ground states found for various combinations. All plots are color-coded according to the overlap of ground states found between the two methods for a given instance, normalized by the degeneracy of the ground state for that instance. Panel (a) shows DW1 ($8$ gauges) \textit{vs } DW1 (another $8$ gauges). Panels (b) and (c) compare the DW1 to SSSV and SQA with parameters optimized as in Fig.~\ref{Albash-fig:DistanceHistogram1}: $(10.56,150k,0.05)$ and $(0.76,10k,0.05)$, respectively. Panel (d) shows the DW1 \textit{vs } SQA with the latter run at a higher temperature ($2.54$mK). Panels (e) and (f) respectively compare SSSV and SQA with the optimized parameters and with SQA at the higher temperature used in (d).}
	\label{Albash-fig:Scatter}
\end{figure} 

We demonstrated in the previous section that using the entire population of energy levels observed, SQA and SSSV correlate better with each other than with the DW1. Yet, when restricted to the ground state success probability, both SQA and SSSV exhibit a strong correlation with the DW1 data \cite{q108,SSSV}. In this section we return to the distribution of ground states and ask whether there is a different measure that does distinguish the DW1, SQA, and SSSV.  To do so we go beyond the success probabilities and compare the actual ground states found by each method. 

\subsection{Correlations colored by overlap}
Consider first the correlation between success probabilities over the set of $1000$ random Ising instances (a measure used in Ref~\cite{q108} to establish that the DW1 and SQA are strongly correlated, while rejecting simulated annealing and classical spin dynamics), plotted in Figs.~\ref{Albash-fig:Overlap-DW1-DW1} and \ref{Albash-fig:Overlap-DW1-SSSV}. Here we see that the SSSV model is as strongly correlated with the DW1 as the DW1 is with itself. To attempt to separate the DW1 and SSSV, we now go a step further and include, as a color scale, the overlap between the sets of ground states found for each instance in $1000$ runs, normalized by the degeneracy $D_i$ of the ground states for that instance. That is, the normalized overlap is defined as 
\beqAlbash
\textrm{overlap}_i^{(X,Y)} = \frac{1}{D_i} G_i^{(X)}\cap G_i^{(Y)}\ ,
\eeqAlbash
where $G_i^{(X)}$ is the set of all ground states found in $1000$ runs of random Ising instance $i$ using method $X$, with $X$ being either one of two sets of $8$ gauges for the DW1, all $16$ gauges for the DW1, or SSSV. We observe, in Fig.~\ref{Albash-fig:Overlap-DW1-DW1} that this quantity is strongly gauge dependent, in that there is no clear separation visible by color. Perhaps unsurprisingly then, there is no clear correlation visible in terms of the normalized overlap when comparing the DW1 and SSSV in Fig.~\ref{Albash-fig:Overlap-DW1-SSSV}. Moreover, the success probability appears to be uncorrelated with the overlap, even for SQA \textit{vs } SSSV, as seen in Fig.~\ref{Albash-fig:Overlap-SSSV-SQA}. Thus, this attempt to distinguish the DW1 and SSSV is inconclusive. 

The situation changes when we consider instead of the success probabilities the fraction of total ground states found for a given instance. This is shown in Figs.~\ref{Albash-fig:Fraction-DW1-DW1}-\ref{Albash-fig:Fraction-SSSV-SQA-beta=3}. In \ref{Albash-fig:Fraction-DW1-DW1}-\ref{Albash-fig:Fraction-DW1-SQA-beta=10} we compare the DW1 to itself, SSSV, and SQA, with the optimized simulation parameters used for the latter two. We observe first of all that the two sets of DW1 gauges are significantly more strongly correlated with each other than SSSV with the DW1, with a Pearson correlation coefficient of $0.903$ for the former [Fig.~\ref{Albash-fig:Fraction-DW1-DW1}] as compared to $0.811$ for the latter [Fig.~\ref{Albash-fig:Fraction-DW1-SSSV}]. 

Figures~\ref{Albash-fig:Fraction-DW1-SSSV} and \ref{Albash-fig:Fraction-DW1-SQA-beta=10} show that the DW1 tends to find a significantly larger fraction of the ground states than both SSSV and SQA. This indicates that the DW1 may be exploring a larger fraction of the ground state manifold than both SQA and SSSV for the chosen parameters, which (as we showed earlier) are the ones that maximize the success probability correlations. We have checked that this conclusion is robust to increasing the number of SSSV sweeps (up to $500k$, not shown).  We have also checked that this conclusion is not sensitive to the particular set of gauge realizations by comparing SSSV with $16$ gauges to SSSV with $16$ different gauges, 
where we observe a correlation coefficient of $0.988$ between the two sets of gauges (not shown). 

This conclusion, however, is not robust to changing the temperature of the simulation. As seen in Fig.~\ref{Albash-fig:Fraction-DW1-SQA-beta=3}, SQA finds a larger fraction of ground states than the DW1 when the SQA simulation temperature is increased to $2.54$mK. This suggests that at the lower temperature that matches the DW1 data, SQA is trapped more often in local minima, which causes it to explore a smaller fraction of the ground state manifold.
More research will be needed to explore the potential of the machine to explore more configurations than the classical algorithms that mimic its behavior,

It is also interesting to compare SSSV to SQA. Fig.~\ref{Albash-fig:Fraction-SSSV-SQA} shows that for the parameter values that best fit the DW1 data, SSSV outperforms SQA for nearly all instances in terms of the fraction of ground states found. However, this conclusion is again a result of the choice of the simulation temperature, since in Fig.~\ref{Albash-fig:Fraction-SSSV-SQA-beta=3} SQA, now at a higher temperature, outperforms SSSV.

We also note that the normalized overlap (color scale) seen in Figs.~\ref{Albash-fig:Fraction-DW1-DW1}-\ref{Albash-fig:Fraction-SSSV-SQA-beta=3} is strongly correlated with the fraction of ground states found by all methods, and they decrease together. That is, when trapping in local minima is more common and fewer distinct ground states are found, all methods explore different regions of the ground state manifold. To some extent the same holds true when comparing different DW1 gauges, as seen in Fig.~\ref{Albash-fig:Fraction-DW1-DW1}, i.e., when trapping occurs different gauges can localize around small and different sets of ground states. 
\subsection{Trace-norm distance between ground subspace distributions}
Keeping in mind the sensitivity of DW1 to gauges and the fact that the number of annealing runs is limited so that instances with a large degeneracy are very likely undersampled, we next restrict our analysis to the instances with a sufficiently small degeneracy. Let $\rho^X$ denote the density matrix of method $X$, $P_0$ the projection onto the ground subspace, and $p_0^X$ the ground subspace (success) probability. Then $\rho_0^X \equiv \frac{1}{p_0^X}P_0 \rho^X P_0$ is a normalized density matrix over the ground subspace, giving each of the degenerate ground states its appropriate weight. We define a ground subspace distance between two methods $X$ and $Y$ as the trace-norm distance between $\rho_0^{X}$ and $\rho_0^{Y}$:
\begin{equation} 
\label{Albash-eqt:distance2}
\mathcal{D}_{\mathrm{GS}} (\rho_0^{X}, \rho_0^{Y}) = \frac{1}{2} \mathrm{Tr} \left|  \rho_0^{X}- \rho_0^{Y}   \right|\ .
\end{equation}
 \begin{figure}[b]
 \centering
	\subfigure[\ DW1 \textit{vs } DW1]{\includegraphics[width=.32\columnwidth]{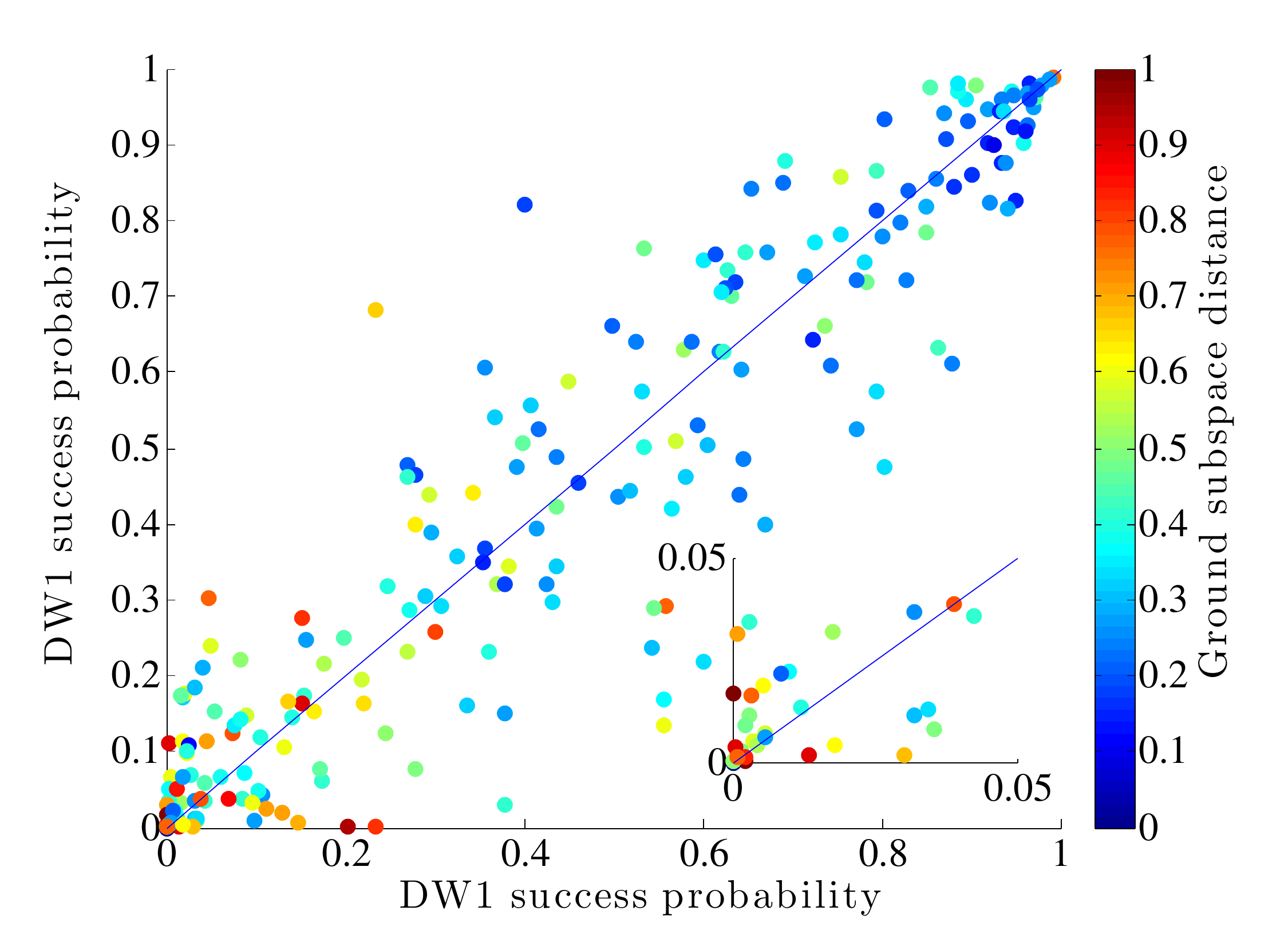}\label{Albash-fig:Scatter1}}
	\subfigure[SSSV \textit{vs } SSSV]{\includegraphics[width=.32\columnwidth]{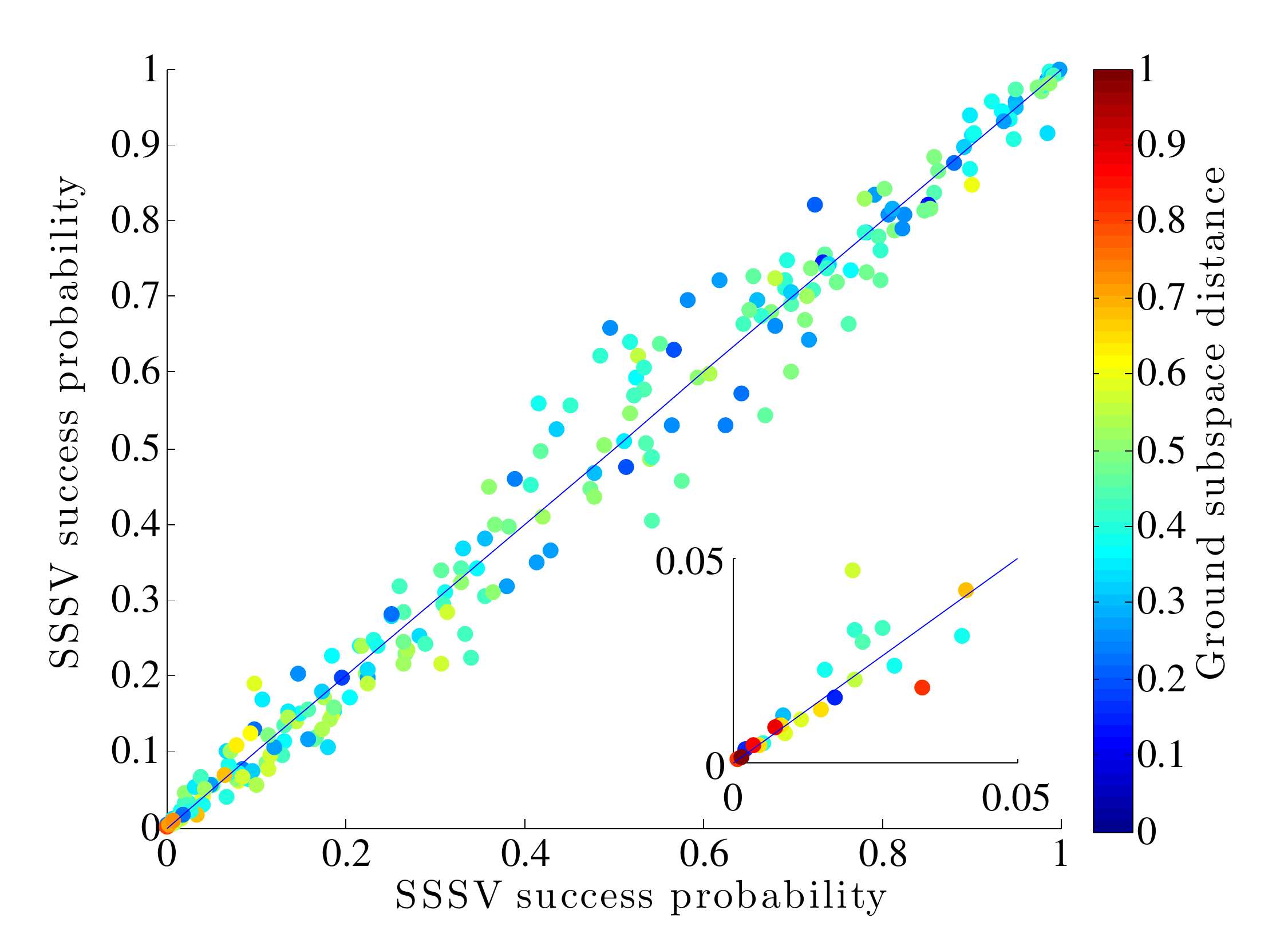}\label{Albash-fig:Scatter5}}
	\subfigure[\ DW1 \textit{vs } SSSV]{\includegraphics[width=.32\columnwidth]{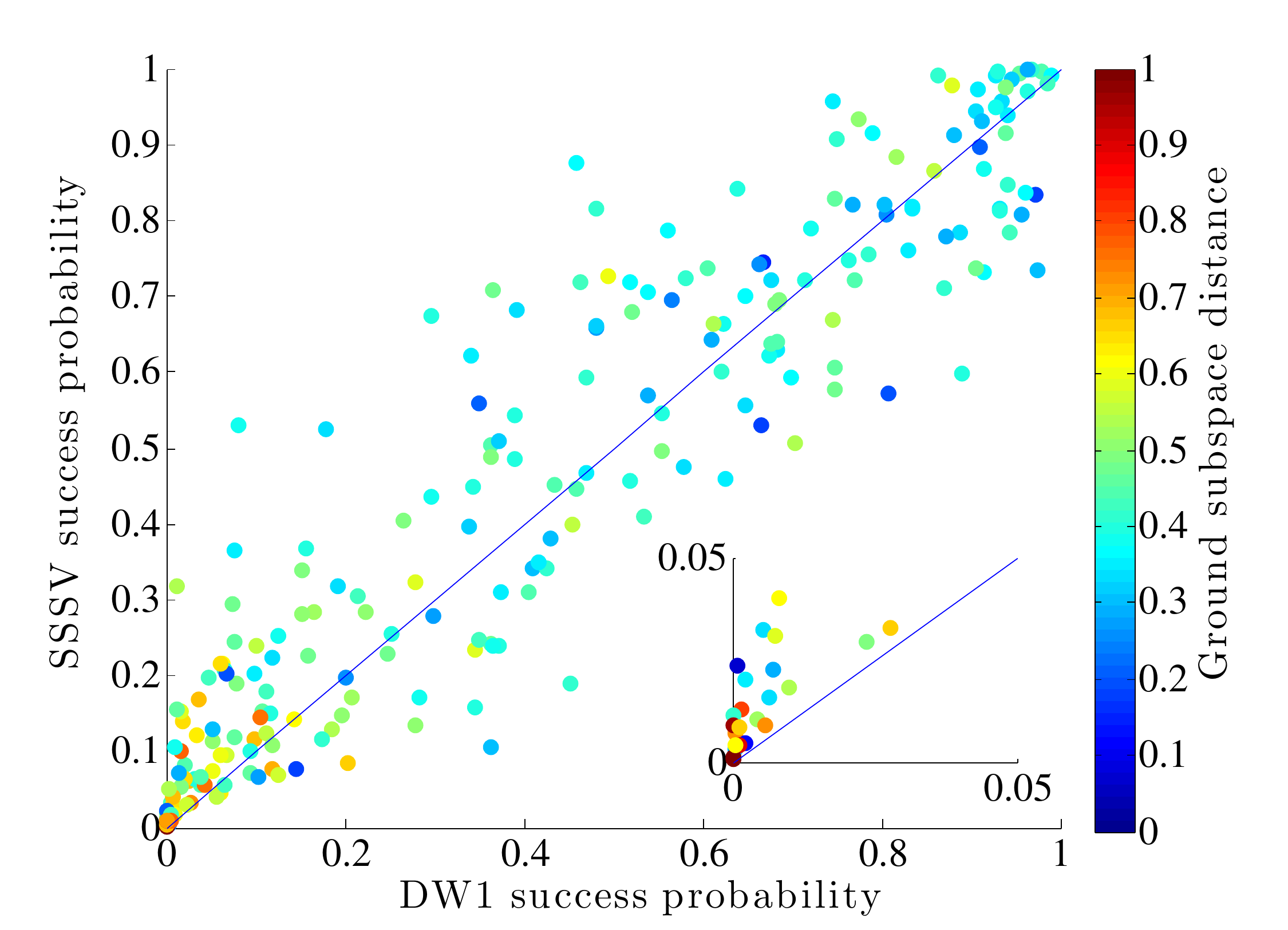}\label{Albash-fig:Scatter2}}
	\subfigure[\ DW1 \textit{vs } SQA]{\includegraphics[width=.32\columnwidth]{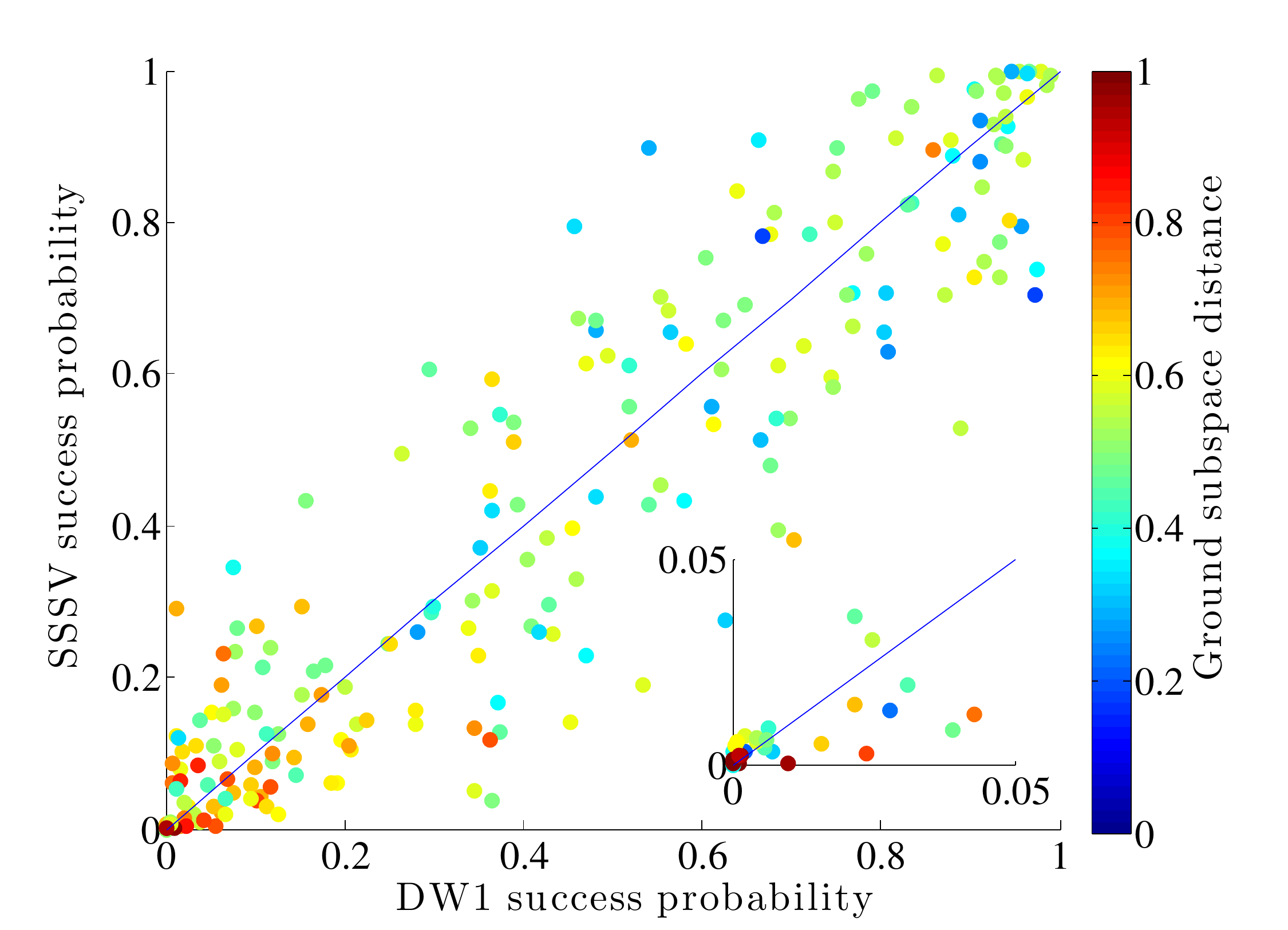}\label{Albash-fig:Scatter3}}
	\subfigure[SQA \textit{vs } SSSV]{\includegraphics[width=.32\columnwidth]{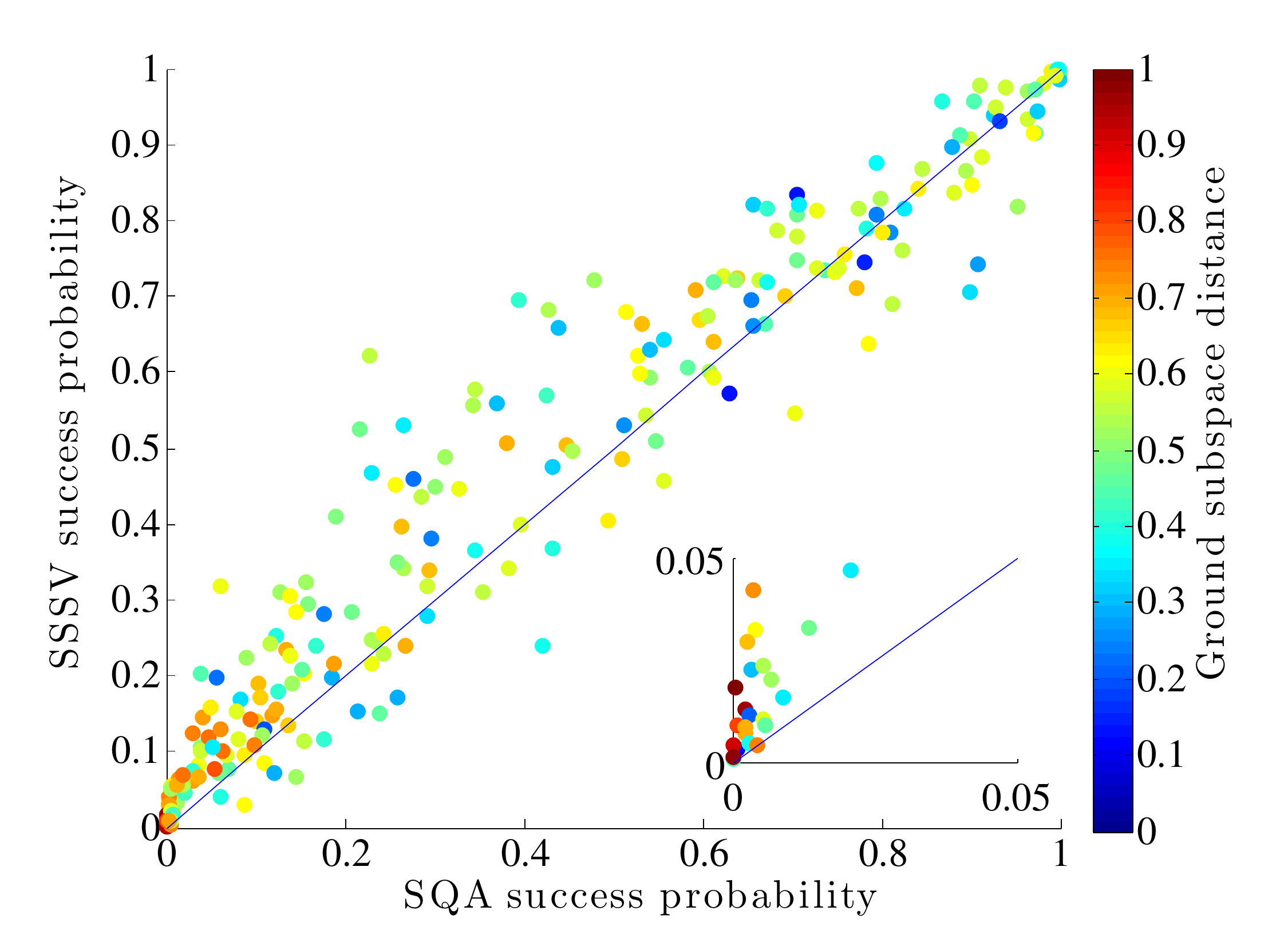}\label{Albash-fig:Scatter4}}
	\caption{Success probability correlation plot for instances with ground state degeneracy of $\leq 100$, color-coded by the ground subspace distance as defined in Eq.~\eqref{Albash-eqt:distance2}. (a) DW1 ($8$ gauges) \textit{vs } DW1 ($8$ other gauges); (b) SSSV ($16$ gauges) \textit{vs } SSSV ($16$ other gauges); (c) DW1 ($16$ gauges) \textit{vs } SSSV; (d) DW1 \textit{vs } SQA; (e) SSSV \textit{vs } SQA.  The insets are zooms of the plots around small success probability. SSSV parameters optimized as in Fig.~\ref{Albash-fig:DistanceHistogram1}: $(10.56,150k,0.05)$.}
	\label{Albash-fig:ScatterDistance}
\end{figure} 

In Fig.~\ref{Albash-fig:ScatterDistance} we plot the correlation of success probabilities over the $1000$ random Ising instances with a degeneracy up to $100$, now color-coded by this ground subspace distance. Figure~\ref{Albash-fig:Scatter1} shows that while for instances with a high success probability this distance is predominantly quite small, it is quite sensitively gauge-dependent, with a sensitivity that increases with problem instance hardness.  This sensitivity to gauges is also present in SSSV as illustrated in Fig.~\ref{Albash-fig:Scatter5} where the distance can be appreciable even when the success probability correlation is high, although the SSSV distance results show little dependence on the success probability.
We observe in Figs.~\ref{Albash-fig:Scatter2} and \ref{Albash-fig:Scatter3} that even though there is a strong success probability correlation between DW1 and SSSV/SQA, the ground subspace distance can be high.  The distance tends to be largest for instances where both solvers have a very low success probability, indicating that for problems that both solvers find hard, different ground states are observed.  However, this is not surprising given that a similar pattern is observed in Fig.~\ref{Albash-fig:Scatter1} when comparing the DW1 against itself. SQA and SSSV appear to be similarly correlated with the DW1, i.e., a similar distribution of distances is visible in Figs.~\ref{Albash-fig:Scatter2} and Fig.~\ref{Albash-fig:Scatter3}. More strikingly, Fig.~\ref{Albash-fig:Scatter4} shows that, in contrast to the comparison to the DW1, the ground subspace distance between SQA and SSSV does not depend as strongly on the success probability. However, for most instances the ground subspace distance between SQA and SSSV is non-negligible ($\gtrsim 0.3$), {indicating that the two methods find different sets of degenerate ground states}, as is also apparent from Fig.~\ref{Albash-fig:Fraction-SSSV-SQA}.

 \begin{figure}[t]
 \centering
	\subfigure[\ SSSV (10.56, 100k, 0.05)]{\includegraphics[width=.32\columnwidth]{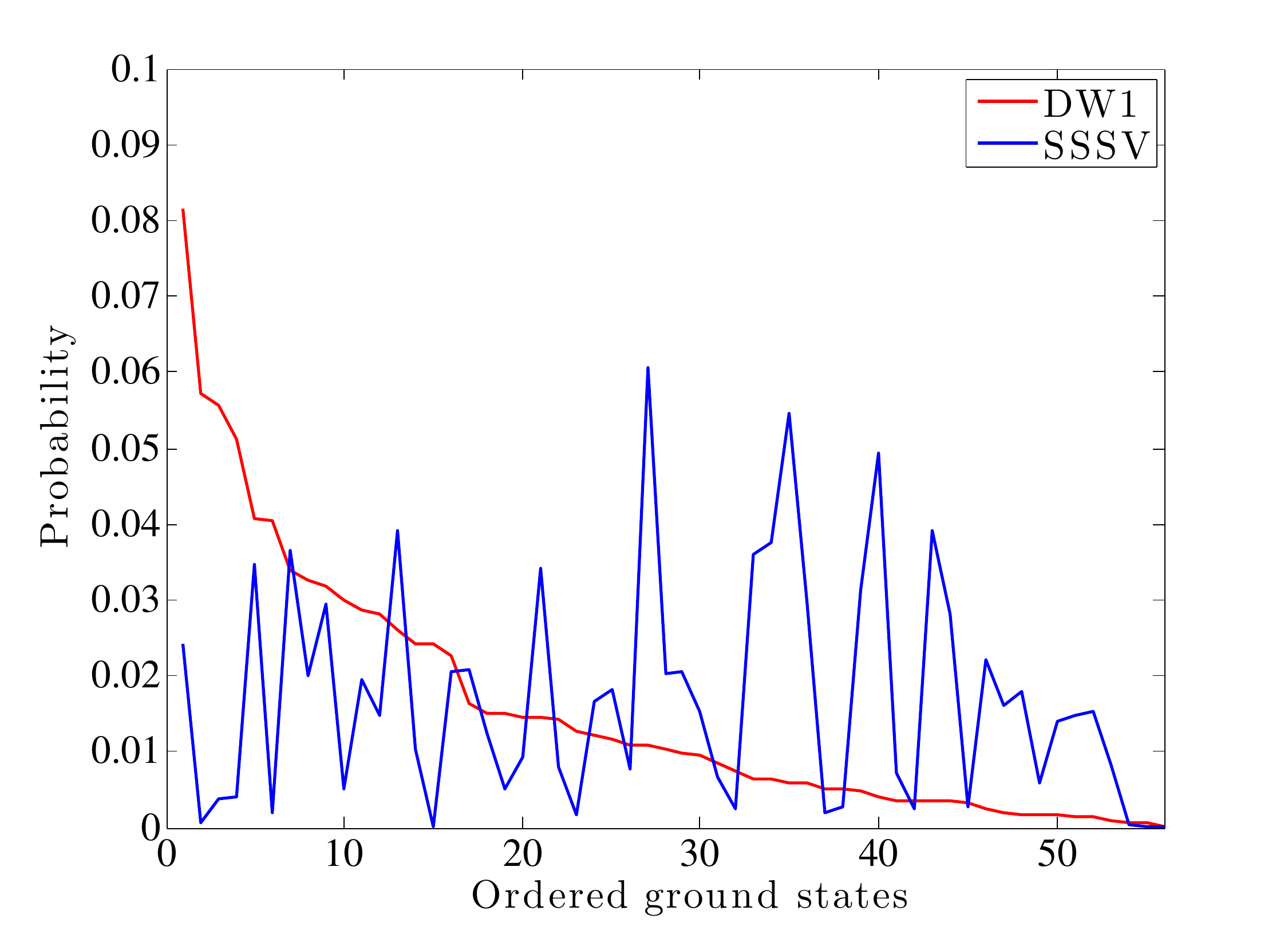}\label{Albash-fig:Instance246_1}}
	\subfigure[\ SSSV (10.56, 150k, 0.05)]{\includegraphics[width=.32\columnwidth]{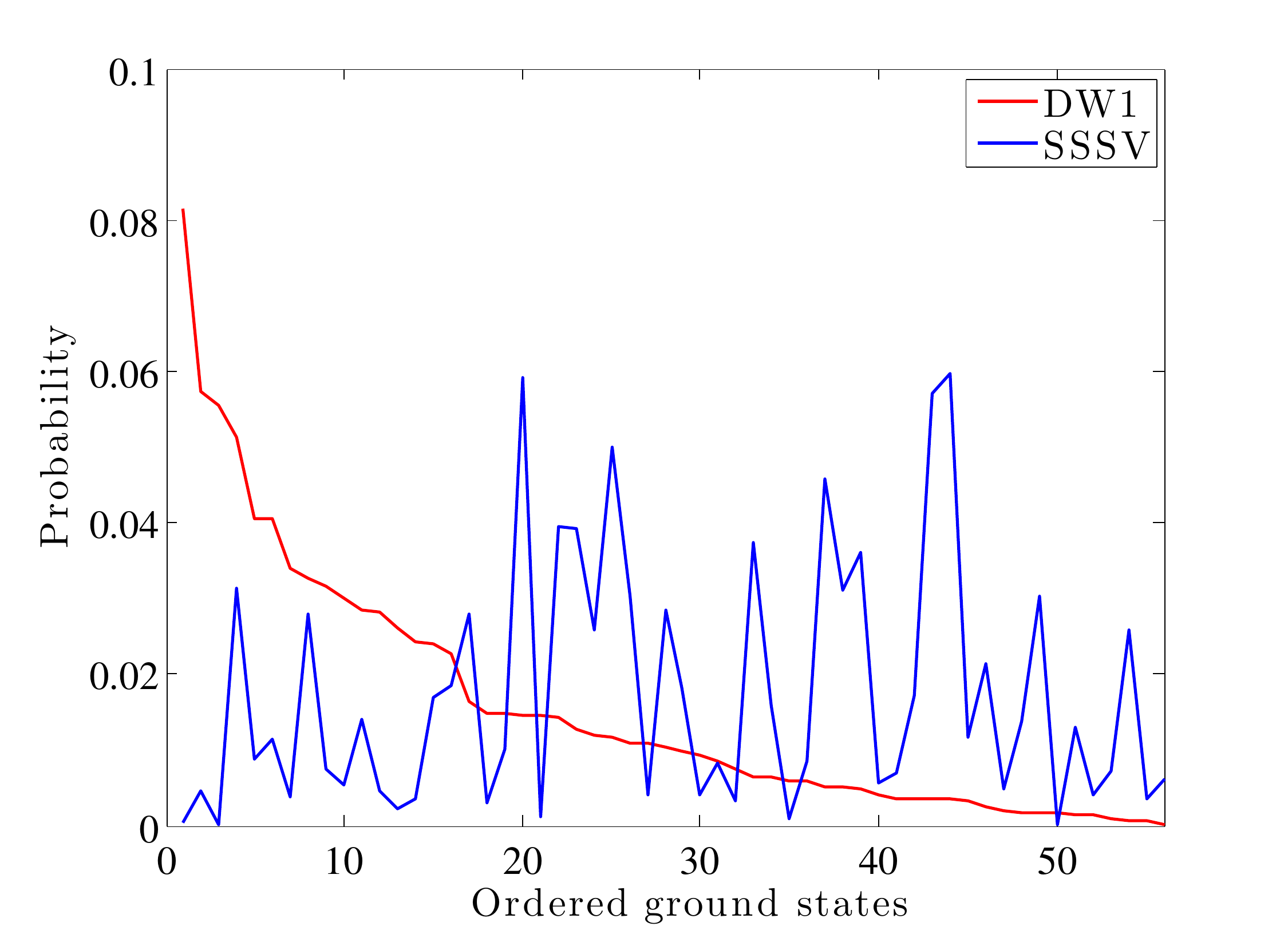}\label{Albash-fig:Instance246_2}}
	\subfigure[\ SSSV (10.56, 150k, 0.075)]{\includegraphics[width=.32\columnwidth]{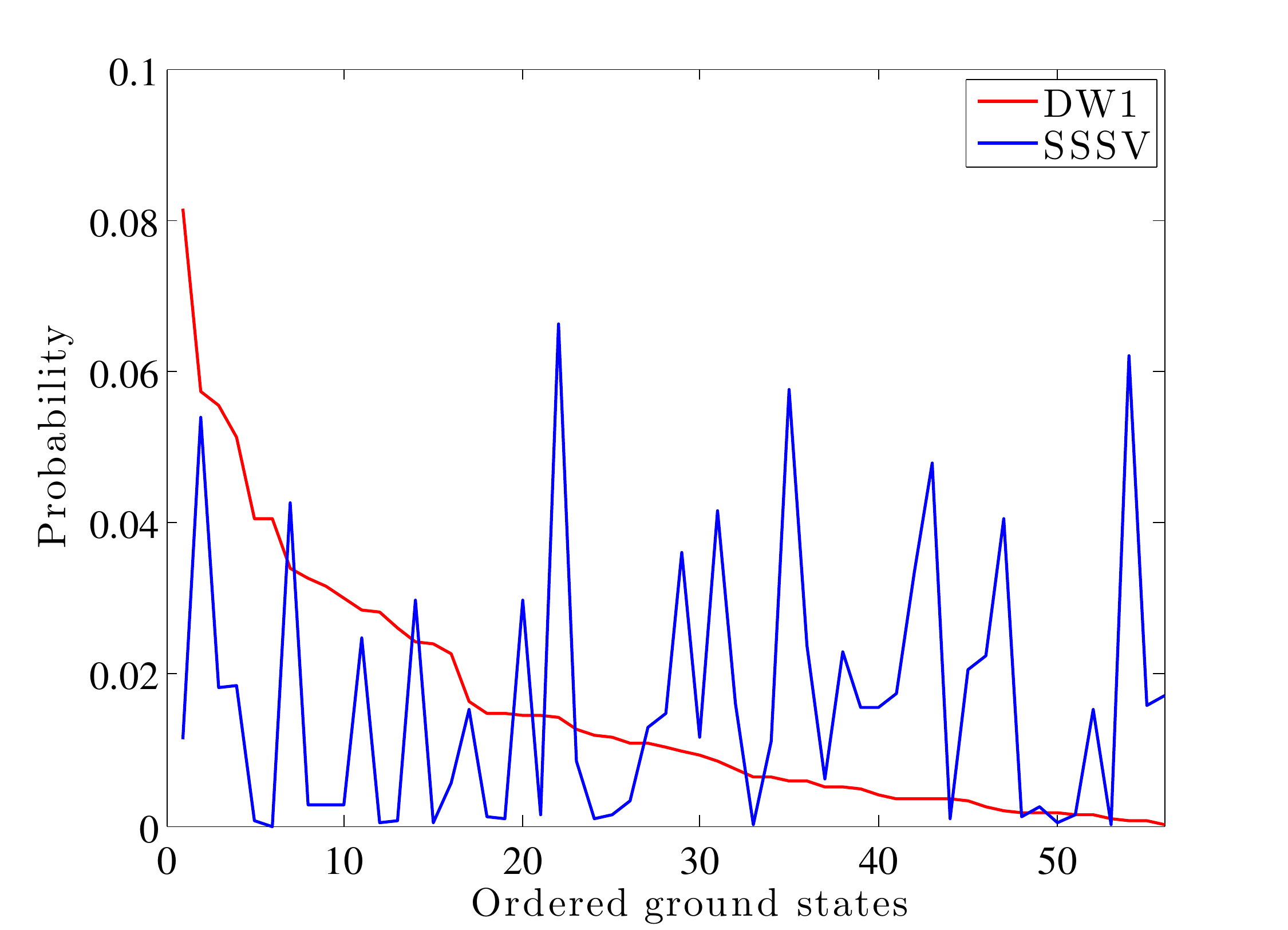}\label{Albash-fig:Instance246_3}}
	\subfigure[\ SSSV (10.56, 100k, 0.05)]{\includegraphics[width=.32\columnwidth]{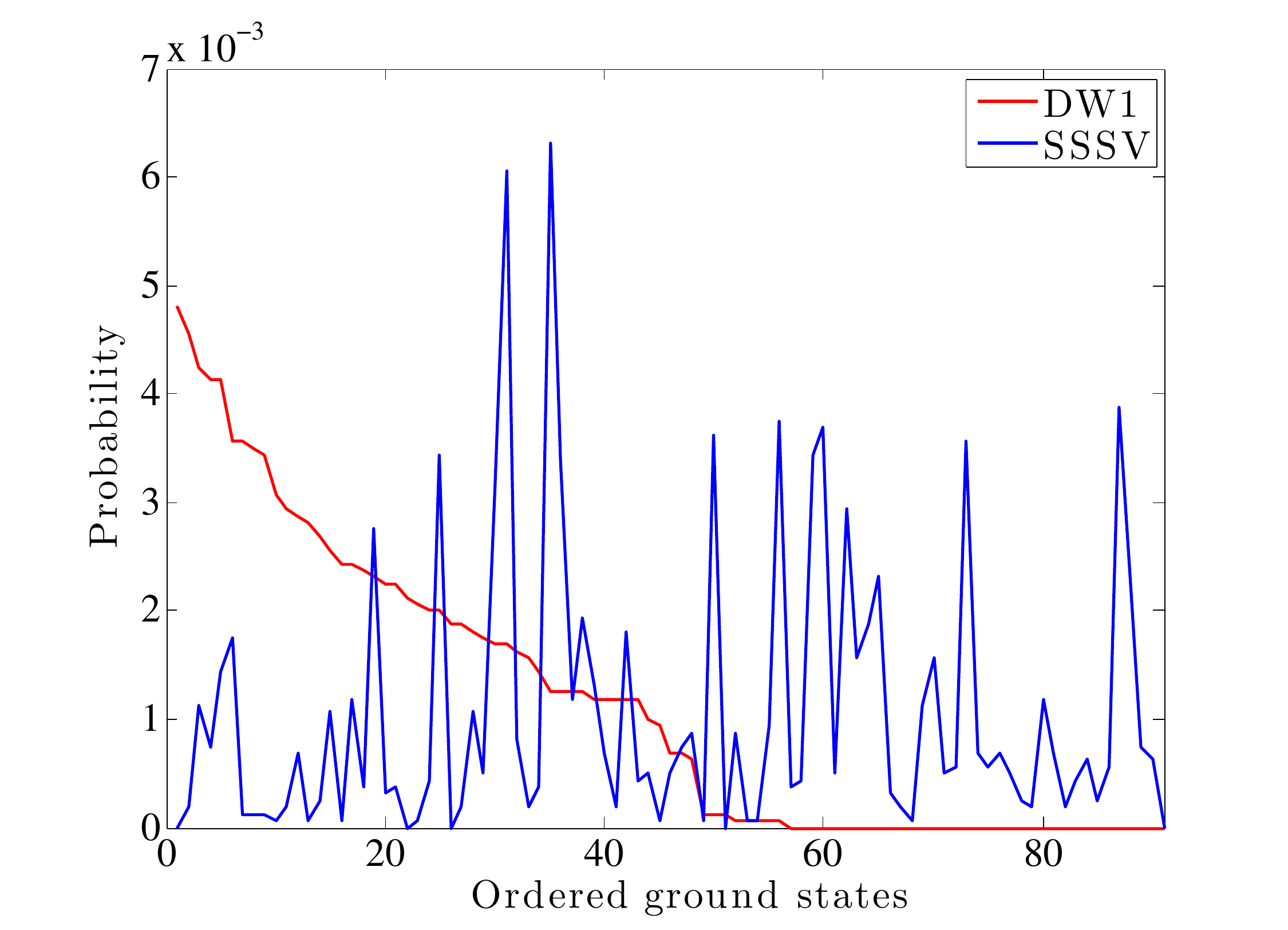}\label{Albash-fig:Instance513_1}}
	\subfigure[\ SSSV (10.56, 150k, 0.05)]{\includegraphics[width=.32\columnwidth]{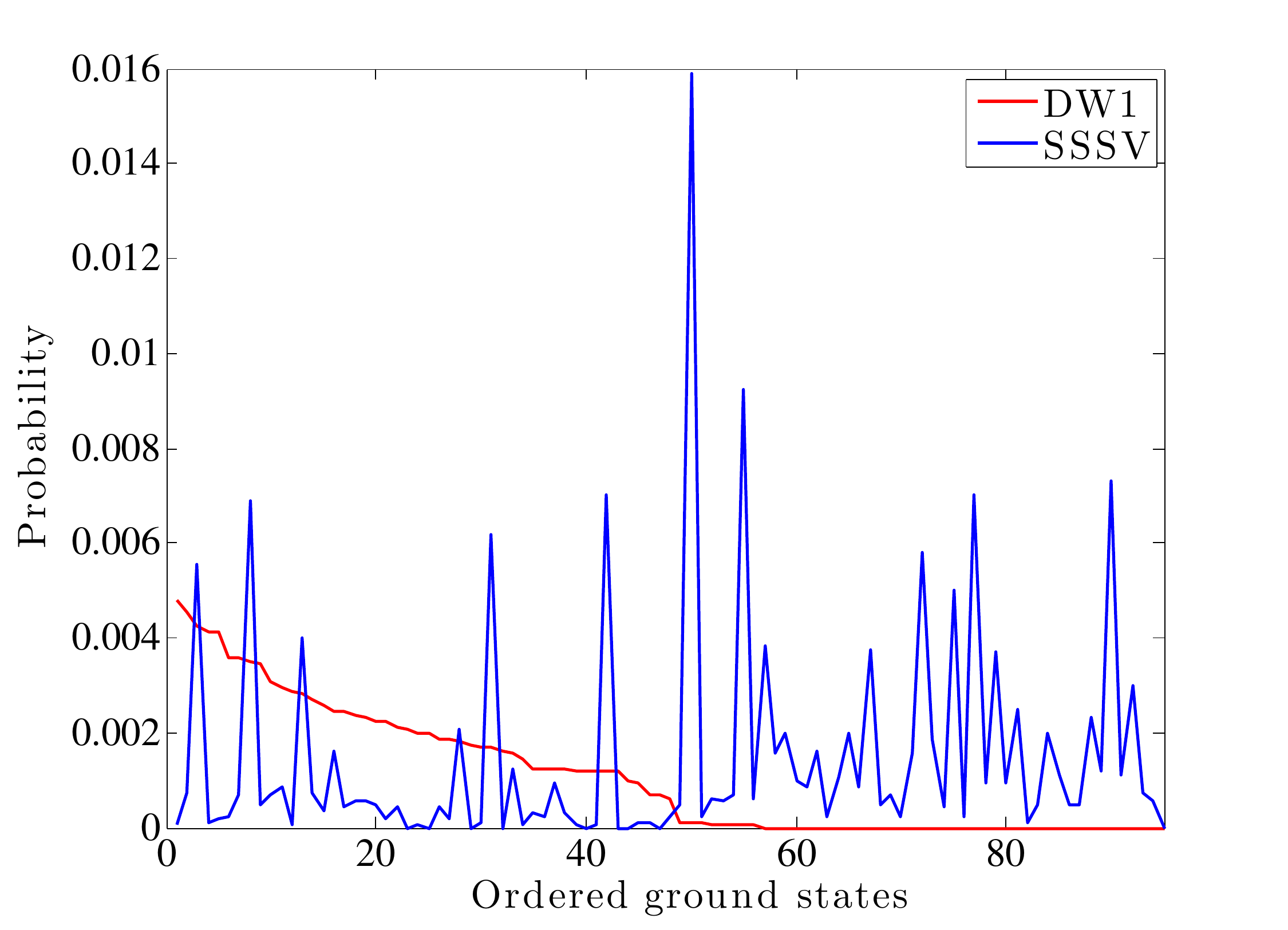}\label{Albash-fig:Instance513_2}}
	\subfigure[\ SSSV (10.56, 150k, 0.075)]{\includegraphics[width=.32\columnwidth]{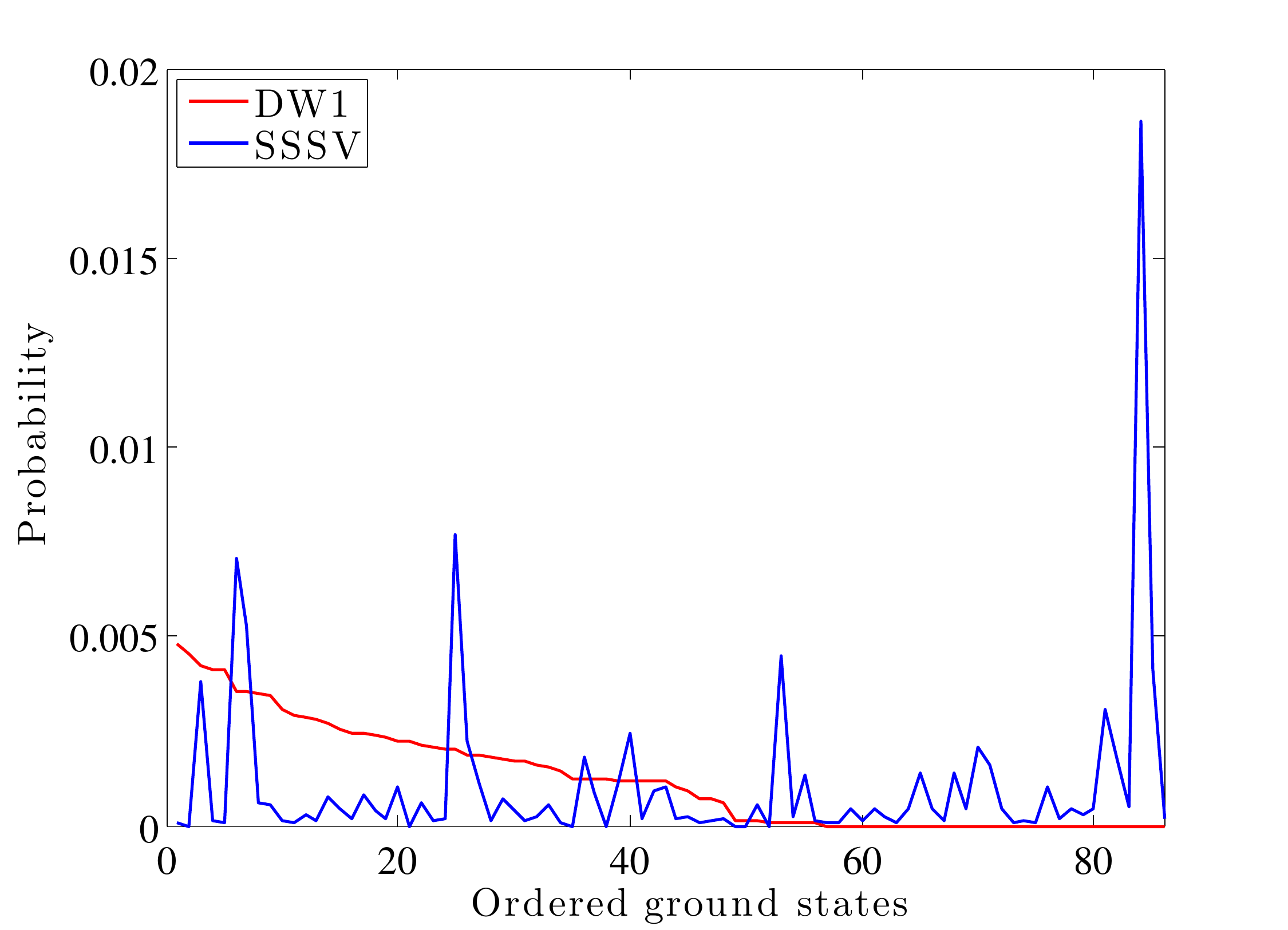}\label{Albash-fig:Instance513_3}}
	\subfigure[\ SQA $(0.76, 10k, 0.05)$]{\includegraphics[width=.32\columnwidth]{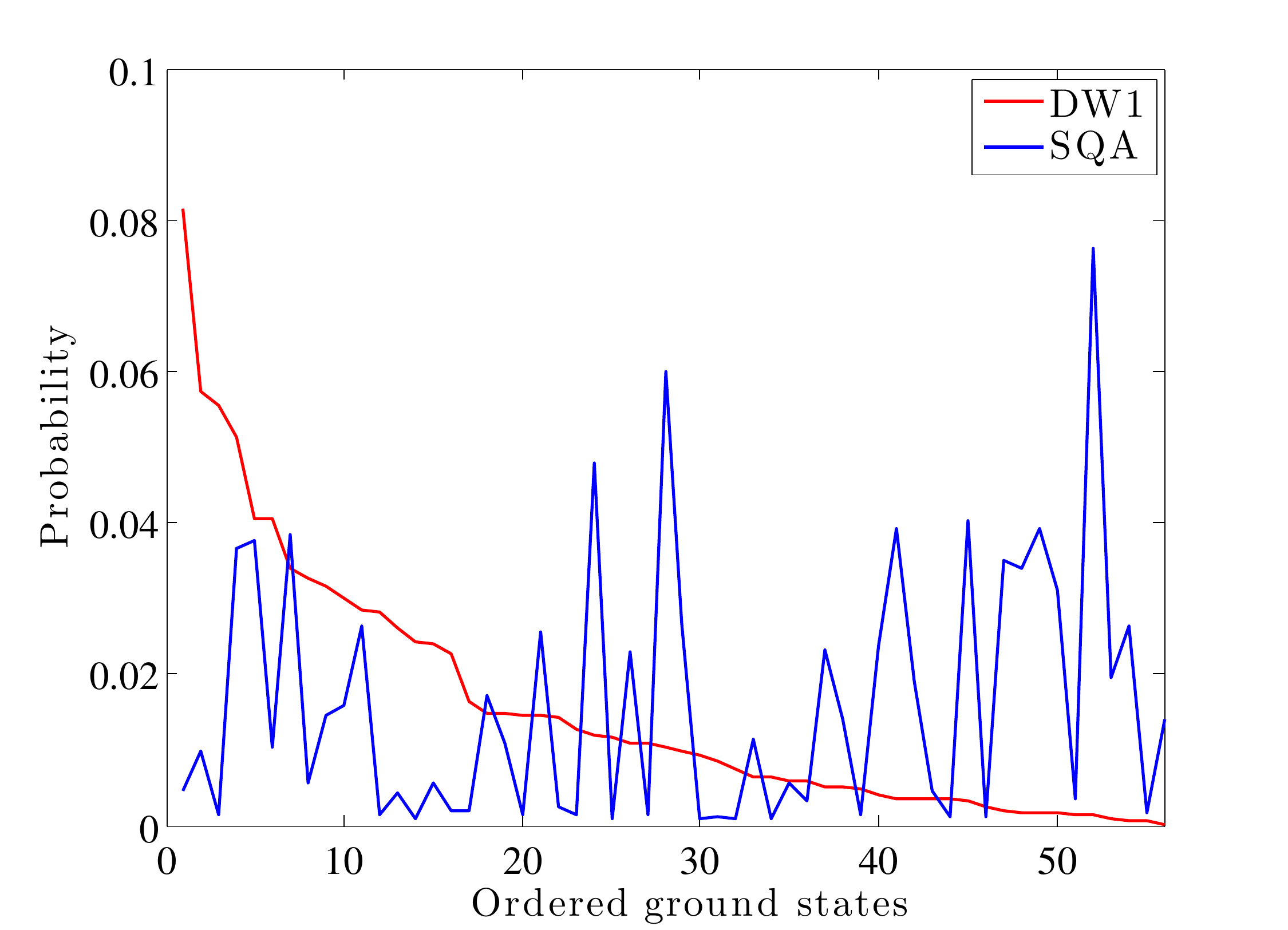}\label{Albash-fig:Instance246_4}}
	\subfigure[\ SQA $(0.76, 10k, 0.05)$]{\includegraphics[width=.32\columnwidth]{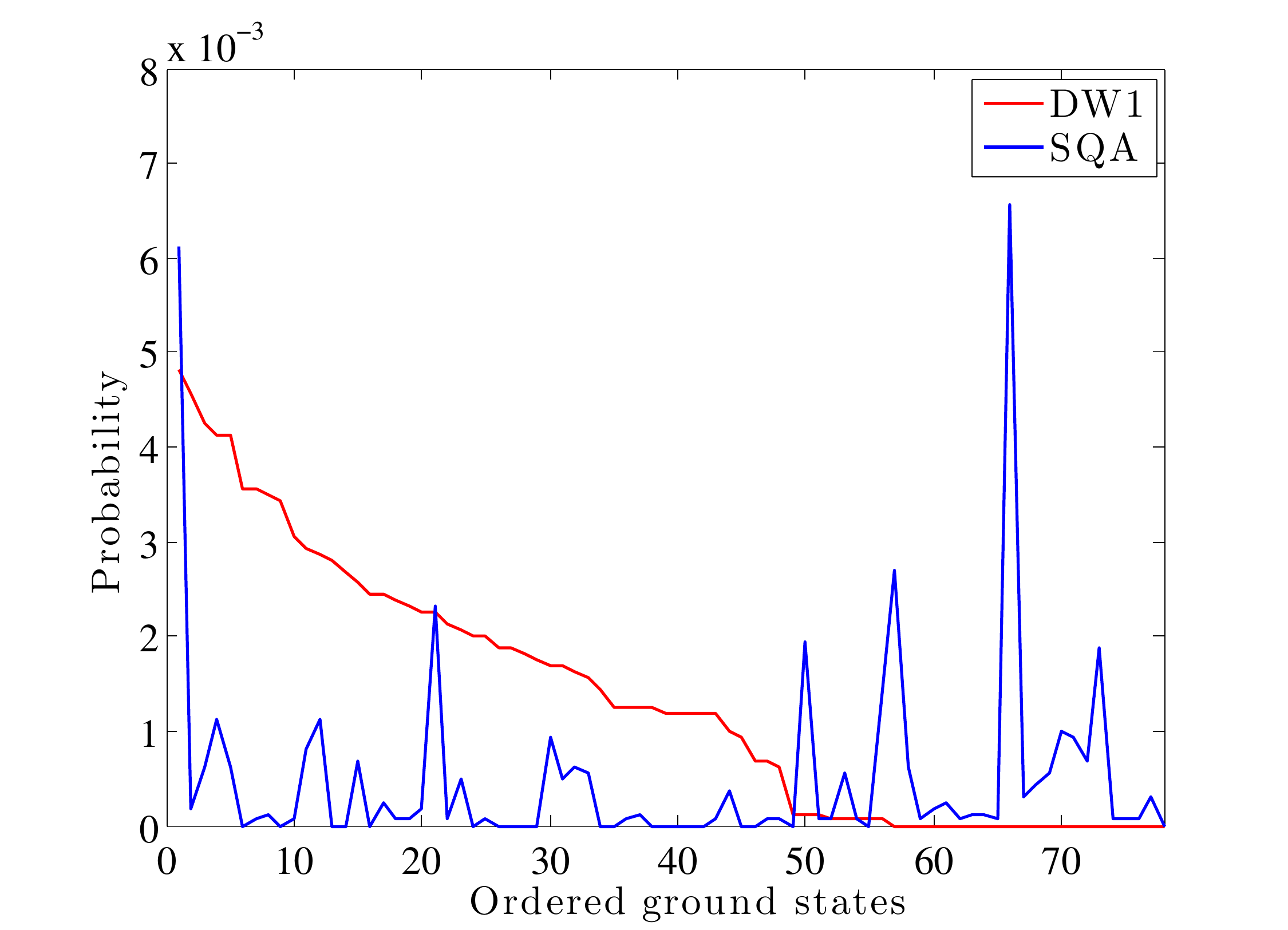}\label{Albash-fig:Instance513_4}}
	\subfigure[\ SSSV $(10.56, 150k, 0.05)$, SQA $(0.76, 10k, 0.05)$]{\includegraphics[width=.32\columnwidth]{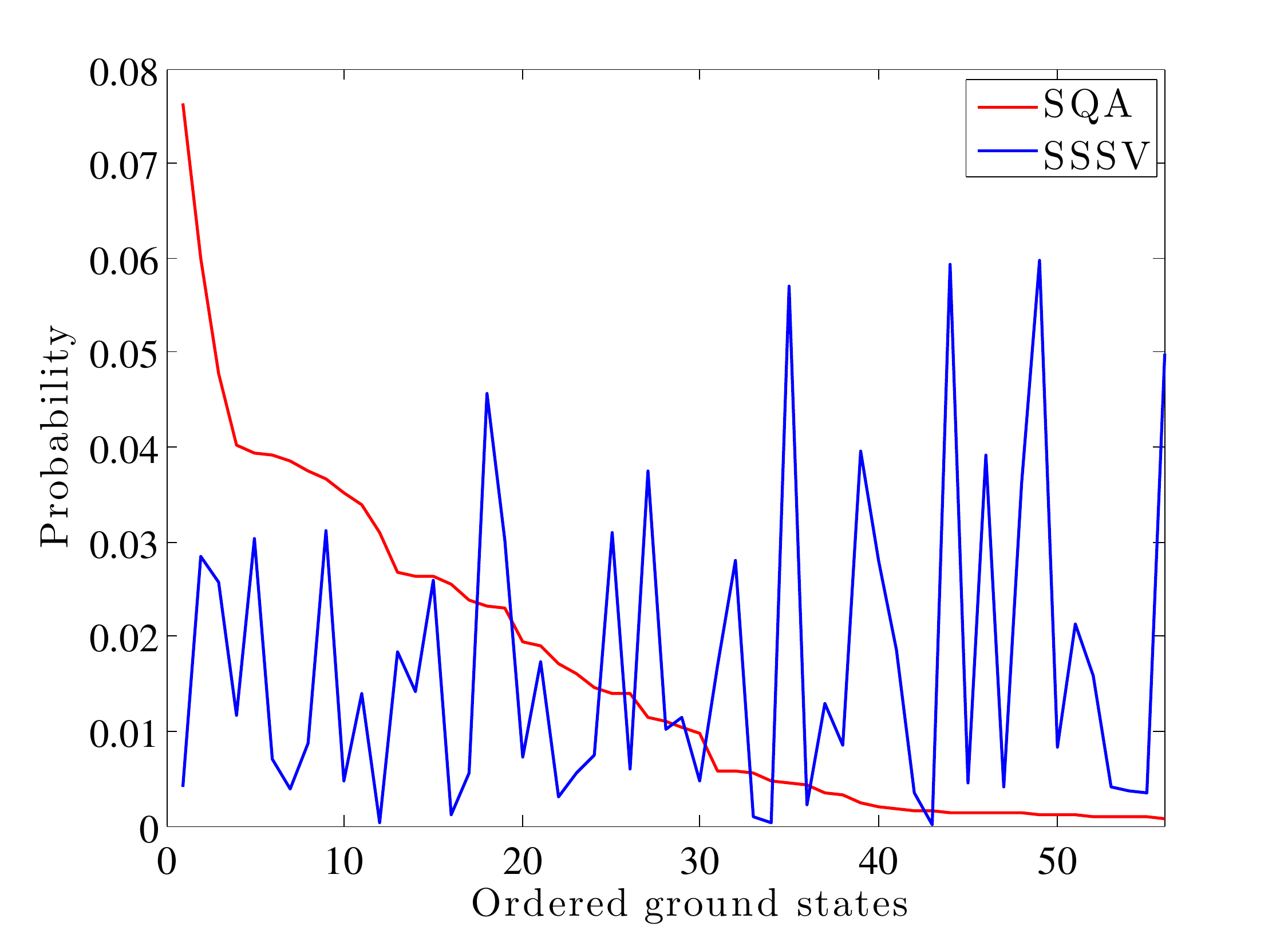}\label{Albash-fig:Instance246_SQA-SSSV}}
	\caption{Ground state probabilities found for an easy and a hard Ising instance from the $1000$ random Ising instances set. The first Ising instance [panels (a)-(c)] is $56$-fold degenerate, the second [panels (d)-(f)] is $96$-fold degenerate.  The ground states are ordered in declining order according to their DW1 probabilities.  The $16$ gauge-averaged DW1 success probability are (a)-(c) $0.878$ (easy) and (d)-(f) $0.106$ (hard), while for SSSV they are (a) $0.963$, (b) $0.978$, (c) $0.948$ (likewise easy), and (d) $0.010$, (e) $0.146$, (f) $0.128$ (likewise hard), due to the use of different SSSV parameters as indicated. Panels (g) and (h) show the same two instances but compare the DW1 to SQA. The corresponding SQA success probabilities are (g) $0.909$ (likewise easy), (h) $0.042$ (likewise hard). Panel (i) shows SQA \textit{vs } SSSV for the same easy instance, sorted by the SQA probabilities.}
	\label{Albash-fig:Instance246_513}
\end{figure} 

Figure~\ref{Albash-fig:ScatterDistance} gives a coarse-grained view of the degeneracy correlations. It is instructive to consider a few specific random Ising instances and compare the ground states found by the different methods. This is the subject of Fig.~\ref{Albash-fig:Instance246_513}, where we plot the probability of each ground state, sorted in decreasing order according to the DW1 results, for two instances, one ``easy"  and one ``hard" (high \textit{vs } low success probability). In the top row [Figs.~\ref{Albash-fig:Instance246_1}-\ref{Albash-fig:Instance246_3}] we compare the DW1 result for the easy instance to SSSV, while varying the SSSV parameters. We do the same for the hard instance in the middle row [Figs.~\ref{Albash-fig:Instance513_1}-\ref{Albash-fig:Instance513_3}]. In the bottom row we compare the DW1 results to SQA for the two instances at the optimal parameters identified earlier. Clearly, the SSSV results are poorly correlated with the DW1, a conclusion that is robust to varying the SSSV parameters. This variation leaves the SSSV model poorly correlated with itself, indicating that the identity of which ground states are found by the SSSV model is not a robust feature, and depends strongly on the number of sweeps and the noise. It is also interesting to note that there are states which are found by the DW1 but not by the SSSV model, and \textit{vice versa}. {This indicates that the two explore different regions of the solution space. For the hard instance it appears that the DW1 explores a smaller region than SSSV, though there too there are solutions that were found by the DW1 but not by the SSSV model.}
We find similar results for SQA [see Figs.~\ref{Albash-fig:Instance246_4} and \ref{Albash-fig:Instance513_4}], although interestingly SQA exhibits a peak at the same position as the DW1 high peak for the hard instance.  Figure~\ref{Albash-fig:Instance246_SQA-SSSV} shows the comparison between SQA and SSSV for the same easy instance; the lack of correlation is apparent, and holds also for the hard instance (not shown). Thus, at last we have found a feature in regards to which SQA and the SSSV model behave differently, at least for the model parameters we have used in our simulations. 

A final caveat is in order: the specific ground states  found are very sensitive to calibration errors of the DW1 device, in the sense that the implemented Hamiltonian may be different from the desired one. In this sense it is difficult to draw definitive conclusions concerning the overlap of ground states  found by each method. A device with smaller calibration errors is clearly desirable in this regard, though the introduction of coupling and local field noise in our simulations mitigates this problem to a large extent, as is indeed evident by the very existence of some overlap between the experimentally observed and simulated ground states.

\section{The SSSV model as a semiclassical limit of the spin-coherent states path integral}
\label{sec:SSSV-deriv}
We have seen that SSSV and SQA agree to a remarkable degree by nearly every measure we have tried. To explain this, we provide a first-principles derivation of the SSSV model in this section, starting from the path integral. 
To connect to the SSSV model our derivation employs spin coherent states, while standard quantum Monte Carlo derivations make use of the computational basis \cite{Suzuki:1976rt}. For this reason we will not find the SSSV model as a classical limit of SQA. However, we will see how the SSSV model can be understood as a certain mean-field approximation to the path-integral, whereby copies of the spin system in the imaginary time direction are made completely decoupled and identical.

Consider the tensor product state of $N$ spin-$\frac{1}{2}$ Bloch coherent states \cite{Arecchi:1972fk}
\begin{equation}
\ket{\Omega} = \bigotimes_{i=1}^N \left( \cos (\theta_i/2) \ket{0}_i + \sin(\theta_i/2) e^{i \phi_i} \ket{1}_i \right) \ ,
\label{Albash-eq:Omegas}
\end{equation}
with $\theta_i \in[0,\pi]$ and $\phi_i \in [0,2\pi)$, i.e., each qubit has a corresponding Bloch sphere vector 
$\vec{v}_i = \left(\sin\theta_i\cos\phi_i,\sin\theta_i\sin\phi_i,\cos\theta_i\right)$.
One can think of the states $\ket{\Omega}$ as describing a collection of coherent single qubits without quantum or classical correlations.  They form an overcomplete spanning set for the $N$-qubit Hilbert space with a completeness relation given by
\begin{equation}
\mathds{1} = \int d \Omega  \ket{\Omega} \bra{ \Omega } \ ,
\end{equation}
where for general spin $S$
\begin{equation}
d \Omega = \prod_i \frac{2 S + 1}{4 \pi} \sin \theta_i d \phi_i d \theta_i \ .
\end{equation}
Let us now consider the partition function of a time-independent Hamiltonian $H$ at some inverse temperature $\beta$:
\begin{equation}
Z = \mathrm{Tr} \left( e^{-\beta H} \right) = \int d \Omega_0 \bra{\Omega_0} e^{-\beta H} |\Omega_0 \rangle \ .
\end{equation}
We can perform an imaginary-time Trotter slicing of $\beta$ and introduce a overcomplete set of spin-coherent states between the Trotter slices:
\begin{equation}
Z =  \lim_{\nu \rightarrow \infty} \int d \Omega_0  \cdots \int d \Omega_{\nu-1} \prod_{n=1}^{\nu} \bra{\Omega_n} e^{-\Delta H }  \ket{\Omega_{n-1}} \ ,
\end{equation}
with $\Delta = \beta/\nu$ and $\ket{\Omega_{\nu}} \equiv \ket{\Omega_0}$. In taking the $\nu \rightarrow \infty$ limit $\beta$ is kept constant and $\Delta$ is sent to zero. Written in this way, we can interpret the index $n$ as labeling a new periodic spatial (imaginary time) direction $\tau$ into which our system is extended.  We can further interpret $\ket{\Omega_n}$ to be the state in the $n$-th slice of this direction, i.e., $\ket{\Omega_n} \equiv \Omega(\tau = n \Delta)$.  

To make progress we would like to Taylor expand $e^{-\Delta H } $ and keep only terms of order $\Delta$. This requires us to make sense of the overlap of neighboring states in the imaginary time direction
\begin{eqnarray}
\braket{\Omega_n|\Omega_{n-1}}  =  \prod_{i=1}  K_{n,i}\ ,
\end{eqnarray}
where
\beqAlbash
K_{n,i} =  \cos \frac{\theta_{n,i}}{2} \cos \frac{\theta_{n-1,i}}{2}  + e^{-i (\phi_{n,i} - \phi_{n-1,i})} \sin \frac{\theta_{n,i}}{2} \sin \frac{\theta_{n-1,i}}{2}\ .
\eeqAlbash
We note that 
\beqAlbash
\left| K_{n,i} \right|^2 = \cos^2(\Theta_{n,i}/2)
\eeqAlbash
where 
\beqAlbash
\cos(\Theta_{n,i}) = \cos\theta_{n-1,i}\cos\theta_{n,i} + \sin\theta_{n-1,i}\sin\theta_{n,i} \cos(\phi_{n-1,i}-\phi_{n,i})\ ,
\eeqAlbash
with $\Theta_{n,i}$ being the angle between the associated Bloch vectors $\vec{v}_{n-1,i},\vec{v}_{n,i}$ \cite{Lieb:1973kx}. Thus $ K_{n,i}$ vanishes only if $\vec{v}_{n-1,i}$ and $\vec{v}_{n,i}$ are anti-parallel ($\Theta_{n,i}=\pi$). Therefore, if we assume the differentiability of the states $\ket{\Omega_n}$ such that $\ket{\Omega_{n-1}} = (1 - \Delta  \partial_\tau) \ket{\Omega_n} + O (\Delta^2)$, i.e., neighboring states in the imaginary time direction differ by an amount of order $\Delta$, then $\braket{\Omega_n|\Omega_{n-1}}$ is never zero and to first order in $\Delta$ we can write
\begin{equation}
\braket{\Omega_n|\Omega_{n-1}} = 1 - \Delta \bra{\Omega_n} \partial_{\tau} \ket{\Omega_n} + O(\Delta^2) \ .
\label{eq:Om-overlap}
\end{equation}
We now expand $e^{-\Delta H}$ and re-exponentiate, which yields, using Eq.~\eqref{eq:Om-overlap}:
\besAlbash
\begin{align}
\bra{\Omega_n} e^{-\Delta H }  \ket{\Omega_{n-1}} &= \bra{\Omega_n} \left( \mathds{1} - \Delta H \right) \ket{\Omega_{n-1}} + O(\Delta^2) \\
&=  \braket{\Omega_n|\Omega_{n-1}} - \Delta  \bra{\Omega_n} H \ket{\Omega_{n-1}} + O(\Delta^2) \\
&= \exp \left[  - \Delta \bra{\Omega_n} \partial_{\tau} \ket{\Omega_n} - \Delta \bra{\Omega_n} H \ket{\Omega_{n}} \right]  + O(\Delta^2)\ .
\end{align}
\eesAlbash
Putting these results together, we have for the partition function:
\besAlbash
\begin{align}
Z   &= \lim_{\nu \rightarrow \infty}   \int d \Omega_0  \cdots \int d \Omega_{\nu}  e^{- \Delta \sum_{n=0}^{\nu-1}  \left[  \bra{\Omega_n} \partial_{\tau} \ket{\Omega_n} +  \bra{\Omega_n} H \ket{\Omega_{n}} \right]} \\
&=  \int \mathcal{D} \Omega \ e^{- \int_0^\beta d \tau  \left[ \bra{\Omega(\tau)} \partial_{\tau} \ket{\Omega(\tau)} + \bra{\Omega(\tau)} H \ket{\Omega(\tau)} \right]} \ ,
\end{align}
\eesAlbash
where periodic boundary conditions are implied, i.e., we set $\Omega(0) = \Omega(\beta)$. However, the term $\bra{\Omega(\tau)} \partial_{\tau} \ket{\Omega(\tau)}$ is a Berry phase and is purely imaginary:
\begin{equation}
\bra{\Omega(\tau)} \partial_{\tau} \ket{\Omega(\tau)} = \frac{i}{2} \sum_{j} \sin^2\left( \frac{\theta_j(\tau)}{2} \right) \dot{\phi}_j(\tau) \ .
\end{equation}
This phenomenon, of having phase factors attached to the classical
Boltzmann weight, is of course a well-known general feature of the mapping between quantum statistical mechanics in $d$ dimensions and classical statistical mechanics in $d + 1$ dimensions (the ``sign problem"), and prevents us from using standard Monte Carlo techniques to estimate the full quantum partition function (see, e.g., Ref.~\cite{Kirchner:2010fr}). It is also clear that the SSSV model does not have a sign problem, so we are led to ask whether there is a limit where we can avoid the Berry phase term in the exponential.  One way to achieve this is to require that the $\phi$'s are constant in the $\tau$ direction.  This then reduces the partition function to 
\begin{equation} \label{Albash-eqt:ZSSSV1}
Z_{\mathrm{SSSV}}  =  \int \mathcal{D} \Omega \ e^{- \int_0^\beta d \tau  \bra{\Omega(\tau)} H \ket{\Omega(\tau)}} \ .
\end{equation}
This limit gives rise to no couplings in the $\tau$ direction, and each $\tau$ slice can be treated separately.  Therefore, using continuity of neighboring states and imposing $\partial_\tau \phi = 0$ should be interpreted as taking a particular classical limit.  Inserting the quantum annealing Hamiltonian \eqref{eq:H}, the classical Hamiltonian given by $\bra{\Omega(\tau)} H \ket{\Omega(\tau)}$ is precisely the SSSV Hamiltonian \eqref{eq:H_SSSV} provided we additionally set $\phi_j=0$ for all $j$. The Metropolis angle updates used in the SSSV model can be understood as estimating the associated partition function.

We note that if we relax the condition of continuity between neighboring states in the imaginary time direction then we must deal with the possibility of $\braket{\Omega_n |\Omega_{n-1}} = 0$ arising from the appearance of anti-parallel Bloch vectors. In this case the derivation presented above does not go through. However, this scenario corresponds to transitions between orthogonal qubit states in successive time-steps, which can be understood as transitions between spin-up and spin-down states, and in this case the standard quantum Monte Carlo derivation in the computational basis is appropriate, which results in the SQA model. It is unclear at this time how to derive the SSSV model in the computational basis setting, rather than using spin-coherent states.

\section{Summary and Conclusions}
\label{sec:conc}
%
In this work we critically reexamined two important  models, SQA and SSSV, of the D-Wave devices. Our motivation for doing so stemmed from seemingly contradictory results, supporting at the same time ``quantum'' and ``classical" explanations of  experiments using the DW1 device  \cite{SSSV,q108}. 

We can summarize our main findings as follows:
\begin{itemize}
\item There is at present no model that completely explains the full set of DW1 random Ising model experiments on $>100$ qubits.
\item SQA and SSSV correctly predict the ground state (``success") probability distribution of the DW1, but do not perfectly describe the observed excited state spectrum or the distribution of ground states.
\item With the exception of the specific set of degenerate ground states found by each method, SQA and SSSV are in strong agreement with each other, indicating that SQA on random Ising instances has an effective classical description within the parameter range of the DW1 experiments. This may also be a consequence of the absence of a finite temperature spin-glass phase in these problem instances \cite{2014Katzgraber}.
\item The SSSV model can be derived as a semi-classical limit of the spin-coherent path integral, which helps to explain why it can closely approximate SQA.
\item The DW1 device found a greater fraction of the total ground state subspace than SSSV and SQA in the parameter regime where both are most strongly correlated with its success probability distribution. While this conclusion may not be robust to a change in SQA or SSSV parameter settings, the set of ground states found by the DW1 does not appear to be strictly contained within the set found by SQA and SSSV.
\end{itemize}

The last conclusion may be encouraging in terms of the potential for using a device such as the DW1 for computational purposes: it complements classical algorithms in terms of the set of ground state solutions found, which can be useful in a variety of applications where one requires not just one but as many valid solutions to an optimization problem as possible. However, the present study does not include simulated annealing and parallel tempering, which are known to find many more ground states than SQA and almost uniformly sample all low lying states \cite{Matsuda:2009uq}.

It is important to acknowledge that our analysis is subject to several potential loopholes. First, we do not know the exact nature of the calibration noise on the DW1 and it is possible that our Gaussian noise model is overly simplistic; there remains the possibility that better noise modeling might improve the observed correlations between the DW1 and SQA or the SSSV model. Second, the annealing schedules shown in Fig.~\ref{Albash-fig:DW1} are not experimentally measured annealing schedules but were calculated using rf-SQUID models with independently calibrated qubit parameters \cite{Trevor}.  Although we found the effect of individual qubit annealing schedules to be very small, a more accurate model of the annealing schedule may also improve the agreement between the DW1 and SQA or SSSV.

We also note that the imperfect correlation of SQA and SSSV with the DW1 excited state spectrum and ground state distribution results is not entirely surprising simply since both SQA and SSSV are phenomenological models. While master equation simulations have been successful in matching ground state probabilities as well as excited states \cite{q-sig,q-sig2}, they are restricted to small system sizes due to their large computational cost, and it remains an open question how to adequately simulate the D-Wave devices at large numbers of qubits.  Interesting new proposals using matrix product states to limit the amount of entanglement in the simulation have been put forth \cite{Crowley:2014qp}, but initial results appear to be inconclusive and are restricted to one-dimensional Ising problems. Thus the quest for efficient and accurate models for quantum annealers remains an important open problem. In particular, the question of whether the DW1 experiments on random Ising instances \cite{q108} make use of large-scale quantum effects remains open, as our work shows that it appears to require models that go beyond both SQA and SSSV.

The fact that  classical algorithms correlate well with the ground state success probabilities suggests that the quantum annealing evolution ``forgets'' its quantum past in the later phases of the evolution.  This might be a symptom of the particular choice of random Ising problems, but might also be a function of the particular annealing schedule used.  Future work will investigate both aspects and elucidate their role in deciding the importance of quantum effects in quantum annealing, and the potential for a quantum speedup \cite{speedup}.
%
\section*{Acknowledgements}
We thank Dr. Zhihui Wang for providing the DW1 data used in Ref.~\cite{q108} and Ze Lei for contributions to the noise model used in this work.  Part of the computing resources were provided by the USC Center for High Performance Computing and Communications.  This research used resources of the Oak Ridge Leadership Computing Facility at the Oak Ridge National Laboratory, which is supported by the Office of Science of the U.S. Department of Energy under Contract No. DE-AC05-00OR22725. The work of T.A. and D.A.L. was supported under ARO MURI Grant No. W911NF-11-1-0268, ARO grant number W911NF-12-1-0523, and the Lockheed Martin Corporation. The work of T.F.R. and M.T. was supported  by Microsoft Research, ERC Advanced Grant SIMCOFE, the Swiss National Science Foundation through the NCCR QSIT. M.T. acknowledges hospitality of the Aspen Center for Physics, supported by NSF grant PHY-1066293.


\end{document}